\PassOptionsToPackage{dvipsnames}{xcolor}
\documentclass[twocolumn, astrosymb]{aastex701}

\usepackage{amssymb, amsmath}
\usepackage{CJKutf8}
\usepackage{bm}
\usepackage{fontawesome5}
\usepackage{academicons}
\usepackage{graphicx}
\usepackage{float}

\shorttitle{Halo assembly history on central stellar mass - richness plane}
\shortauthors{Xu et al.}

\graphicspath{{./}{figures/}}

\hypersetup{colorlinks=true,
            citecolor=MidnightBlue,
            linkcolor=MidnightBlue,
            filecolor=magenta,
            urlcolor=MidnightBlue}
\defcitealias{Xu2024paper1}{Xu+25}

\def\apjs{{ApJS}}

\def\photoi{\texttt{Photutils.isophote}}

\def\h{\hskip -3 mm}

\def\deg{$^{\circ}$}

\def\lax{{$\mathrel{\hbox{\rlap{\hbox{\lower4pt\hbox{${\sim}$}}}\hbox{$<$}}}$}}
\def\gax{{$\mathrel{\hbox{\rlap{\hbox{\lower4pt\hbox{${\sim}$}}}\hbox{$>$}}}$}}
\def\simlt{\lower.5ex\hbox{$\; \buildrel < \over {\sim} \;$}}
\def\simgt{\lower.5ex\hbox{$\; \buildrel > \over {\sim} \;$}}

\def\logm10{{$\log (M_{\star,10\ \mathrm{kpc}}/M_{\odot})$}}
\def\logm30{{$\log (M_{\star,30\ \mathrm{kpc}}/M_{\odot})$}}
\def\logm50{{$\log (M_{\star,50\ \mathrm{kpc}}/M_{\odot})$}}
\def\logm100{{$\log (M_{\star,100\ \mathrm{kpc}}/M_{\odot})$}}

\def\mhalo{{$M_{\mathrm{halo}}$}}

\def\mh200b{{$M_{\mathrm{200b}}$}}
\def\mh200c{{$M_{\mathrm{200c}}$}}

\def\exsitu{{\textit{ex situ}}}

\def\mdpl2{\texttt{MDPL2}}

\def\illustris{{\tt Illustris}}
\def\tng{{\tt IllustrisTNG}}

\def\dsigma{{$\Delta\Sigma$}}

\definecolor{LightGray}{gray}{0.85}
\definecolor{Tab1}{RGB}{114, 158, 206}
\definecolor{Tab2}{RGB}{255, 158,  74}
\definecolor{Tab3}{RGB}{103, 191,  92}
\definecolor{Tab4}{RGB}{174, 199, 232}
\definecolor{Tab5}{RGB}{255, 187, 120}
\definecolor{Tab6}{RGB}{152, 223, 138}
\definecolor{Tab7}{RGB}{255, 152, 150}
\definecolor{Tab8}{RGB}{197, 176, 213}
\definecolor{hpurple}{HTML}{7E16DF}

\newcommand{\code}[1]{\textbf{\texttt{#1}}} 
\newcommand{\temp}[1]{\textcolor{OliveGreen}{#1}} 
\newcommand{\change}[1]{\textcolor{Black}{#1}} 

\begin{document}
\begin{CJK*}{UTF8}{gbsn}

\title{The Connection between Halo Assembly History and the Stellar Distribution I: The Central Stellar Mass - Richness Plane}

\correspondingauthor{Song Huang}
\email{shuang@tsinghua.edu.cn, xus21@mails.tsinghua.edu.cn}

\author[0000-0002-4460-0409]{Shuo Xu (许朔)}
\affiliation{Department of Astronomy, Tsinghua University, Beijing 100084, China}
\email{xus21@mails.tsinghua.edu.cn}

\author[0000-0003-1385-7591]{Song Huang (黄崧)}
\affiliation{Department of Astronomy, Tsinghua University, Beijing 100084, China}
\email{shuang@tsinghua.edu.cn}

\author[0000-0002-3677-3617]{Alexie Leauthaud}
\affiliation{Department of Astronomy and Astrophysics, UCO/Lick Observatory, University of California, 1156 High Street, Santa Cruz, CA 95064, USA}
\email{alexie@ucsc.edu}

\author[0000-0001-9568-7287]{Benedikt Diemer}
\affiliation{Department of Astronomy, University of Maryland, College Park, MD 20742, USA}
\email{diemer@umd.edu}

\author[0009-0003-8358-8320]{Katya Leidig}
\affiliation{Department of Astronomy, University of Maryland, College Park, MD 20742, USA}
\email{kleidig@umd.edu}

\author[0000-0002-2897-6326]{Conghao Zhou (周丛浩)}
\affiliation{Physics Department, University of California, Santa Cruz, CA 95064, USA}
\affiliation{Santa Cruz Institute for Particle Physics, Santa Cruz, CA 95064, USA}
\email{czhou64@ucsc.edu}

\author[0000-0003-3843-7366]{Carlo Cannarozzo}
\affiliation{New York University Abu Dhabi, PO Box 129188, Abu Dhabi, United Arab Emirates}
\affiliation{Center for Astrophysics and Space Science (CASS), New York University Abu Dhabi}
\email{carlo.cannarozzo@nyu.edu}

\begin{abstract}

    Recent observations suggest intrinsic links between the stellar components of massive halos and their assembly histories. Using massive halos with $\log[M_{\rm halo}/M_\odot] \geq 13.0$ from the \texttt{IllustrisTNG-300} simulation, we explore differences in secondary halo properties between samples selected by central galaxy stellar mass and satellite richness. For group-like masses, we find that halos selected by the outskirts stellar mass of their central galaxies between 50 and 100 kpc ($M_{\star,[50,100]}$) have similar halo mass distributions to those selected by intrinsic satellite richness ($\lambda_{10, R_{200}}$), but differ significantly in assembly history: $M_{\star,[50,100]}$-selected halos are more concentrated and formed earlier. At higher masses ($\sim10^{14}~M_\odot$), intrinsic richness becomes the better halo-mass proxy, but only when projection effects are ignored. \change{We find that the difference-set test selects comparable halo-mass samples most effectively when the two proxies have similar scatter}. Projection effects and baryonic physics do not alter the main trends but remain important caveats for observational applications. Our findings show that the central stellar mass-richness plane is a powerful tool for selecting halos with particular accretion histories, offering new insights into the galaxy-halo connection in the era of deep imaging and precision lensing surveys such as Euclid and LSST.
    
\end{abstract}



\section{Introduction} 
    \label{sec:intro}
    
    Massive dark matter halos in the present-day universe (e.g., $\log[M_{\rm 200c}/M_\odot]\gtrsim13.0$) form from the highest peaks of the primordial density field (\citealt{Press1974ApJ}) and, at low redshift, host galaxy groups and clusters with a massive central galaxy. Their abundance, clustering, and assembly history encode rich cosmological information while shaping the galaxy populations they contain (e.g., \citealt{Bocquet2019ApJ}; \citealt{Moster2018MNRAS}; \citealt{Behroozi2019MNRAS}). To exploit these halos for cosmology, one needs a halo-mass proxy that is readily observable, has low intrinsic scatter, and can be calibrated through weak gravitational lensing (e.g., \citealt{Mandelbaum2006MNRAS}; \citealt{Murata2018ApJ}). With ongoing and upcoming deep imaging and spectroscopic surveys (e.g., Hyper Suprime-Cam (HSC), Euclid, and the Dark Energy Spectroscopic Instrument (DESI), the two most accessible optical proxies are satellite richness and the stellar mass of the central galaxy, both of which we focus on in this work.

    Richness, which directly traces subhalo abundance, is an established proxy (e.g., \citealt{Andreon2010MNRAS}; \citealt{Murata2018ApJ}) used in cluster finders such as redMaPPer (\citealt{Rykoff2014ApJ}) and CAMIRA (\citealt{Oguri2014MNRAS}). However, richness-based proxies suffer from serious systematic biases: in particular, line-of-sight structure (the projection effect) can bias both the richness measurement and the lensing signal (e.g., \citealt{Sunayama2020MNRAS}; \citealt{Wu2022MNRAS}; \citealt{Zhou2024PhRvD}).

    The central galaxy, surrounded by an extended stellar halo or intracluster/intra-group light (ICL/IGrL; e.g., \citealt{Contini2021Galax}; \citealt{Montes2022NatAs}), offers a complementary proxy. Crucially, its performance depends sensitively on the physical scale at which the stellar mass is measured: both observations and hydrodynamical simulations show that the outskirt stellar mass (e.g., $M_{\star,[50,100]~{\rm kpc}}$) traces halo mass better than the aperture mass within the galaxy (e.g., \citealt{Huang2022}; \citealt{Xu2024paper1}, hereafter \citetalias{Xu2024paper1}; \citealt{Zhou2025JCAP}). This outskirts component is dominated by \exsitu{} stars accreted through mergers with satellites (e.g., \citealt{RodriguezGomez2016MNRAS}; \citealt{Cannarozzo2023MNRAS}) and thus reflects the host halo's cumulative accretion, but its faintness makes it difficult to measure, requiring deep imaging and careful background subtraction (e.g., \citealt{Huang2018}).

    Both proxies exhibit significant scatter with halo mass. Beyond observational uncertainties (e.g., \citealt{Rozo2011ApJ}), part of this scatter is intrinsic (e.g., \citealt{Bradshaw2020MNRAS}; \citealt{Chen2024ApJ}), indicating that secondary halo properties such as assembly history also shape galaxy observables. On the one hand, this can introduce selection bias when building large halo samples from these proxies; on the other hand, it offers a valuable opportunity to probe how halo assembly history imprints itself on the stellar content of massive halos. Understanding this connection is therefore crucial for both cosmology and the physics of massive galaxy formation.

    Recent studies have begun to map this connection. In cosmological simulations such as \texttt{The300} and \tng{}, the stellar-mass fractions of different halo components have been shown to be tightly coupled to assembly history, with earlier-forming halos showing higher ICL (or BCG+ICL) fractions but lower satellite mass fractions (e.g., \citealt{Contreras-Santos2024A&AICL}; \citealt{Montenegro-Taborda2025MNRAS}). Observationally, the central stellar mass--richness plane has proven a powerful diagnostic: \citet{Zu2021MNRAS} found that, at fixed richness, halos with more massive central galaxies tend to be more concentrated. Most such observational results, however, rest on richness-selected samples, which may carry their own selection biases and provide an incomplete view of the galaxy--halo connection.

    In this work, we instead use a sample of massive halos from the \tng{}-300 hydrodynamical simulation to study the connection among the satellite population, the central galaxy, and the halo assembly history in a more comprehensive and unbiased way. We develop a \emph{difference-set test} to compare halo properties across samples selected by different halo-mass proxies, the outskirts stellar mass of the central galaxy versus satellite richness, using directly observable stacked lensing profiles, and we trace the physical origin of the differences through halo assembly histories and secondary halo properties. We also assess potential systematics such as projection effects and baryonic physics. In the future, we will extend this work to the central galaxy's stellar profile.

    The paper is organized as follows.
    Section \ref{sec:data} introduces the hydro-simulation {\tt TNG300} and the associated simulated data used in this work.
    Section \ref{sec:methods} describes our method for measuring central galaxy stellar mass and satellite richness of simulated massive halos, as well as the method of extracting their lensing profile and halo mass accretion history.
    Our main findings are presented in Section \ref{sec:results}, followed by detailed discussions in Section \ref{sec:discussions} about the physical implications of our findings and potential caveats.

    The \tng{} Project adopted the $\Lambda$CDM model with parameters $\Omega_m=0.3089$,~$\Omega_{\Lambda}=0.6911$,~$\Omega_b=0.0486$,~$H_0=67.74$~km~s$^{-1}$ Mpc$^{-1}$ and a \citet{Chabrier2003PASP} initial mass function.
    For dark matter halo mass, we adopt $M_{200c}$ -- the spherical overdensity mass with a radius of 200 times the critical density of the universe for all simulations.
 

\section{Simulation Data} 
    \label{sec:data}

    \tng{} (e.g., \citealt{Pillepich2018MNRAS}; \citealt{Springel2018MNRAS}; \citealt{Nelson2018MNRAS}; \citealt{Naiman2018MNRAS}; \citealt{Marinacci2018MNRAS}) is a suite of cosmological large-scale magnetohydrodynamical (MHD) simulations with an improved treatment of gas cooling and photo-ionization, star formation and the ISM, stellar evolution and feedback, and SMBH feedback (e.g., \citealt{Pillepich2018aMNRAS}), with respect to the original \illustris{}\footnote{\url{https://www.illustris-project.org/}} project (\citealt{Vogelsberger2014MNRAS}).  It adopts a new kinetic SMBH feedback mode and improves the model of the galactic wind with many other changes (\citealt{Weinberger2017MNRAS}) compared to the \illustris{}. The \tng{} simulations provide three simulation volumes with different dark-matter and baryonic mass resolutions. This work uses the high-resolution simulation with the largest volume (box size 302.6 cMpc): {\tt TNG300-1}. This simulation has a baryon particle resolution of $1.1\times10^7~M_\odot$ and a softening length of 1480 pc. 

    With its large volume, {\tt TNG300-1} can sample the high-mass end of the halo and stellar mass function reasonably well, providing 3,388 halos with $\log[M_{200{\rm c}}/M_\odot]>13$ at $z=0.4$\footnote{We choose $z=0.4$ to be consistent with the mean redshift of the massive galaxy sample used by previous works using the HSC data (e.g., \citealt{Huang2018}). Using $z=0$ or any lower-redshift snapshot should not alter the main conclusion of this work.} 
    

\subsection{Subhalo Catalog and Halo Merger Trees}
    \label{ssec:shcat_trees}

    In the \tng{} simulations, most particles are assigned to subhalos using the \texttt{SUBFIND} algorithm \citep{Subfind2001MNRAS} following the application of the Friends-of-Friends (FoF) algorithm \citep{FoF1985ApJ}. Using these particles, we compute each subhalo's physical properties and access them via its {\tt SubhaloID}. Note that we count all satellites above a $M_{\star}$ threshold within $R_{200\rm c}$, not just the satellites from the FoF group. In addition to the true richness within the $R_{200\rm c}$ sphere, we also count satellites within a projected 2D cylinder for comparison with observations affected by the projection effect.
    
    In addition to subhalo properties, understanding halo formation and growth requires knowledge of their hierarchical assembly history. Halo/subhalo merger trees in cosmological simulations provide detailed records of how smaller halos or galaxies merge over time to form larger ones. In this work, we adopt the {\tt SubLink} merger trees \citep{RodriguezGomez2015MNRAS} from {\tt TNG300-1} to extract the mass accretion histories of the massive halos in our sample.
    
\subsection{Halo Concentration, Accretion Rate, and Mass Accretion History}
    \label{ssec:conc}

    In simulations, the systematic variations of secondary halo properties at fixed halo mass, such as the halo concentration, halo spin, and accretion rate, are physically connected to the halo assembly histories (e.g., \citealt{Wechsler2002ApJconc}; \citealt{Wang2020MNRAS}; \citealt{DOnghia2007MNRASspin}; \citealt{Chen2020ApJ}), which is a strong motivation to infer them in observations (e.g., \citealt{Giocoli2014MNRAS}; \citealt{Gilman2020MNRAS}; \citealt{Xu2021ApJconc}). In this work, we use halo concentration and accretion rate as tracers of halo assembly history, adopting values from \citet{Anbajagane2022}. We define concentration, as $c_{200c}=R_{200c}/r_s,$ where $r_s$ is the scale radius of the Navarro-Frenk-White (NFW) profile (\citealt{NFW1997ApJ}). We define the mass accretion rate as in \citet{Diemer2017ApJSSPATA}, namely, over a single dynamical timescale to reduce noise.
    
    \begin{equation}
        \Gamma_{\rm dyn}(t)=\frac{\ln[M(t)]-\ln[M(t-t_{\rm dyn})]}{\ln[a(t)]-\ln[a(t-t_{\rm dyn})]},
    \end{equation}
    where $M=M_{\rm 200m}$ is the spherical overdensity mass with a radius of 200 times the mean density of the universe, as opposed to $M_{200c}$.

    Beyond this instantaneous rate, we follow the full mass accretion history of each halo along the main branch of its {\tt SubLink} merger tree (\S\ref{ssec:shcat_trees}), constructed at the subhalo (galaxy) level. At each node, we read the halo mass $M_{200c}$ ({\tt Group\_M\_Crit200} in {\tt TNG}), centered on the host FoF group's potential minimum. To suppress fluctuations during mergers, we adopt the peak halo mass history $M_{\rm peak}(z)$ (e.g., \citealt{Hearin2021OJAp}), the maximum mass the first progenitor has attained up to redshift $z$, and stack it across the halos of interest, bootstrapping for uncertainties. Owing to FoF limitations, we verify at each snapshot that the first progenitor is the central galaxy of its FoF group, excluding the 7 halos that fail this condition along the main branch.

\subsection{Dark Matter Only Simulation}
    \label{ssec:DMO}

    To explore the impact of baryonic effects on lensing profiles, we match massive halos between {\tt TNG300}-1 and its dark-matter-only (DMO) version, {\tt TNG300}-1-Dark. We also consider massive subhalo positions to improve matching accuracy when a massive halo in {\tt TNG300} is classified as a subhalo in the DMO simulation. This DMO simulation shares the same initial conditions as its baryonic counterpart and is consistent with it on large, linear scales. However, hydrodynamical simulations include baryonic processes such as cooling, star formation, and feedback, leading to significant differences between the two approaches on small scales (e.g., \citealt{Anbajagane2022}; \citealt{Beltz-Mohrmann2021ApJ}; \citealt{Duffy2010MNRAS}).
    

\section{Methods} 
    \label{sec:methods}

    In this section, we first describe measurements of stellar mass (\S\ref{ssec:Mstarmea}) and richness (\S\ref{ssec:richmea}) for the simulated massive galaxies. Then, we focus on the dark matter halos themselves by introducing the measurement of the weak lensing profiles (\S\ref{ssec:lensmea}). Finally, we present the methodology for exploring the connections between observables and underlying halo properties under fixed halo-mass distributions (\S\ref{ssec:test}).

    \begin{figure*}[p]
    \centering
    \includegraphics[width=1.1\linewidth,trim=3cm 0 0 3cm]{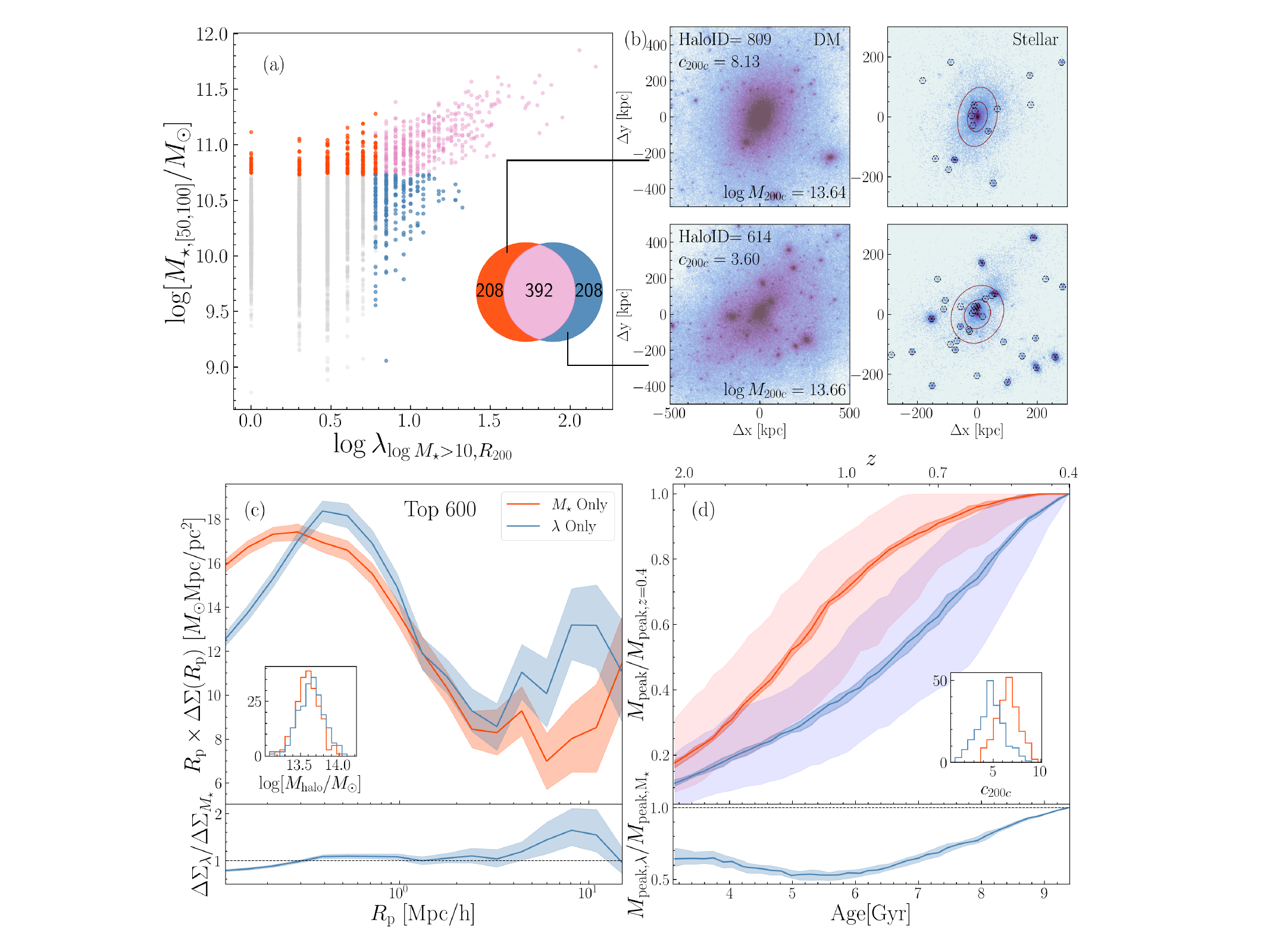}
    \caption{
        Comparison of the lensing profile and halo assembly history between two subsets of massive halos in the TNG300 simulation, derived from a ``difference-set" test on the richness ($\lambda_{10, R_{200}}$) - outskirt stellar mass of the central galaxy ($M_{\star,[50,100]}$) plane. \\
        {\bf Panel (a)}: We get these two subsets by taking the difference sets of two TopN ($N=600$) samples separately based on richness and outskirt mass and exhibit this selection on the $\lambda_{10, R_{200}}$ - $M_{\star,[50,100]}$ plane. \change{There are $208$ halos only selected by richness ($\lambda$ Only, Red Points) or outskirt mass ($M_{\star}$ Only, Blue Points) and $392$ halos selected by both proxies (Pink Points)}\\
        {\bf Panel (b)}: Dark matter and stellar mass maps of two example halos with similar mass but different concentrations.  
        {\it Left Column}: The dark matter mass maps with 1 Mpc$\times$1 Mpc size. Their HaloID, mass, and concentration are noted. 
        {\it Right Column}: The stellar mass maps in the central regions (600 kpc$\times$600 kpc) of the two halos. The red annulus illustrates the measurement of outskirt stellar mass $M_{\star,[50,100]}$ while the black dashed circles show some of the satellites' positions for the measurement of satellite richness $\lambda_{8.5}$. This panel illustrates our scenario: halos with higher concentrations (formed earlier) tend to have fewer satellites but more prominent outskirts with higher stellar mass.\\
        {\bf Panel (c)}: The stacked lensing profile with $R_p\in[0.1,17]{~\rm Mpc/h}$ of $\lambda$-only (blue) and $M_\star$-only (red) samples (upper panel) and their ratio (bottom panel). The shaded areas show the bootstrapped error. $M_\star$-only halos show a stronger lensing signal in the central region than $\lambda$-only halos, indicative of their higher concentration on average. In the outer region, the lensing profile of $M_\star$-only halos is smaller than that of $\lambda$-only halos. The inner panel shows their similar halo mass distribution.\\  
        {\bf Panel (d)}: The normalized peak halo mass history from $z=2$ to $z=0.4$ of these two samples and their ratio. The dark-shaded region in the upper panel represents the error in the average halo assembly history, while the shaded region in the bottom panel indicates the error in the ratio, both calculated via bootstrapping. The light-shaded region in the upper panel displays the $1\sigma$ range of the distribution of the halo assembly history. This panel indicates that $M_\star$-only halos form earlier than $\lambda$-only halos. The inner panel shows their different concentration distribution.\\
        The \texttt{Jupyter} notebook for reproducing this figure can be found here: \href{https://github.com/Xuchuyi/cenvssat/blob/main/FigureNotebooks/Fig1acd.ipynb}{\faGithub}.
        }
    \label{fig:figkey}
    \end{figure*}

    \subsection{Aperture \& Outskirt Stellar Mass}
        \label{ssec:Mstarmea}

    Central galaxies in massive halos are predominantly elliptical with smooth stellar distributions, motivating a 2-D stellar mass profile approach (e.g., \citealt{Ardila2021MNRAS}; \citealt{Cannarozzo2023MNRAS}; \citealt{Xu2024paper1}) to facilitate comparison with observations. We project 3-D stellar particles onto 300 × 300-pixel maps (1 kpc/pixel) along the XY direction to create 2-D stellar mass maps, including only stellar particles from the central region of the Friend-of-Friends (FoF) group while excluding satellite contribution.

    Following \citet{Huang2018} and \citet{Li2022}, we then perform 1-D isophote analysis using the {\tt isophote} module in \photoi{} (as detailed in \citetalias{Xu2024paper1}) to define approximate elliptical isophotes from stellar mass maps. This yields mean values of ellipticity and position angle that characterize the average isophotal shape.
    
    Our simulation-based stellar mass maps intrinsically exclude satellites and contaminants, eliminating the need for object masking \footnote{\citet{Leidig2025arXiv} showed that satellite galaxies contribute approximately half of the local stellar mass density at $\sim100$ kpc from the central galaxy.}.
    Because ellipsoidal isophotal fitting dominates the shape recovery, this halo-finder-based subtraction is as effective as the satellite masking used in \citet{Leidig2025arXiv}, \change{who applied an observational masking procedure to the TNG simulation and obtained measurements of the diffuse stellar component consistent with \citetalias{Xu2024paper1}. This agreement shows that our measurement can be reproduced by the masking procedures developed for observational data (e.g., \citealt{Kluge2020ApJS}; \citealt{Montenegro-Taborda2025MNRASb}).}
    We directly measure aperture and outskirt stellar masses by summing enclosed stellar particles within elliptical apertures defined by these average isophotes (illustrated by the red annulus in Fig. \ref{fig:figkey}b).

    In this work, we define an aperture or outskirt stellar mass in kpc (absolute physical size), which facilitates straightforward application to observational data and improves comparability with observations. Specifically, we utilize 3 apertures defined at 30, 50, and 100 kpc to measure the stellar mass of central galaxies. 
    We used fixed-size apertures rather than effective-radius-defined apertures, since the latter are subject to systematic photometric uncertainties. \footnote{\change{We have verified that effective-radius-based stellar mass, e.g., $M_{\star,[4,8]R_{\rm e}}$, yields consistent results in both the Top-$N$ test and the difference-set test.}}
    In this work, the stellar mass enclosed within the 30 kpc aperture ($M_{\star,30~{\rm kpc}}$) approximately corresponds to the stellar mass of the brightest central galaxy (see also \citealt{Pillepich2018MNRAS}). The 100 kpc, on the other hand, corresponds to the approximate detection limit of the extended stellar halo in \citet{Huang2018}. In addition, we use the outskirts mass between 50 kpc and 100 kpc ($M_{\star,[50,100]~{\rm kpc}}$) to trace the stellar mass associated with the extended stellar halo. 

    \subsection{Richness Measurement}
        \label{ssec:richmea}

        We measure richness by counting satellite subhalos that pass both a stellar-mass and a spatial cut, adopting two choices for each. For the stellar-mass threshold, we use $10^{8.5}~M_\odot$ (about 30 times the baryon-particle mass of {\tt TNG300-1}, i.e., the smallest resolved galaxies) and $10^{10}~M_\odot$ (roughly the $0.2~L_\star$ limit used for satellites by the redMapper algorithm). For the spatial cut, we count satellites either within a sphere of radius $R_{200c}$, which approximates the true richness, or within a cylinder of radius $R_{200c}$ and various half-lengths aligned with the three box axes, which captures the projection effect (e.g., \citealt{Sunayama2020MNRAS}; \citealt{Lee2025PhysRevD}); the second subscript records this choice (e.g., $\lambda_{10, R_{200}}$). Table \ref{tab:richness} lists the resulting definitions.

        We deliberately avoid star formation rate or color cuts to prevent systematic biases from the \tng{} physics model, and we also record the summed satellite stellar mass as an alternative richness-like metric. In what follows, we focus on the $10^{10}~M_\odot$ threshold, which most closely matches observational richness; Appendix \ref{app:choiceRich} justifies this choice.

    \begin{table*}[tb]
        \caption{Different richness definitions and their symbols}
        \centering
        \begin{tabular}{|c|c|}
        \hline
        \textbf{Richness} & \textbf{Definition} \\ \hline
            $\lambda_{10, R_{\rm 200}}$ & \parbox{12cm}{Number of satellites with stellar mass $> 10^{10}~M_\odot$, within a spherical region centered on the central galaxy with radius $R_{\rm 200c}$} \\ \hline
            $\lambda_{8.5,R_{\rm 200}}$ & \parbox{12cm}{Similar to $\lambda_{10,R_{\rm 200}}$, but with stellar mass cut at $10^{8.5}~M_\odot$} \\ \hline
            $\lambda_{10,10}$ & \parbox{12cm}{Number of satellites with stellar mass $> 10^{10}~M_\odot$, within a cylindrical region centered on the central galaxy with radius $R_{\rm 200c}$ and half-length 10 Mpc} \\ \hline
            $\lambda_{10,40}$ & \parbox{12cm}{Similar to $\lambda_{10,10}$, but with cylindrical region half-length 40 Mpc} \\ \hline
        \end{tabular}
        \label{tab:richness}
    \end{table*}

    \subsection{Weak Lensing Measurements in Simulation}
        \label{ssec:lensmea}

        The excess surface density profile, or \dsigma{} profile, is the most commonly used weak-lensing measurement today. For massive halos, the \dsigma{} profiles not only reflect their total mass distributions but also encode hints related to their secondary halo properties (e.g., \citealt{Xhakaj2022MNRAS}). Here, we define the \dsigma{} profiles as
        $$
            \Delta\Sigma(R)\equiv\bar{\Sigma}(<R)-\bar{\Sigma}(R) \,,
        $$
        where $\bar{\Sigma}(R)$ denotes the azimuthally averaged mass surface density at $R$ and  $\bar{\Sigma}(<R)$ means the average mass surface density enclosed within $R$.     

        Observationally, galaxy–galaxy lensing infers mass profiles from the tangential shear $\gamma_t$ of background galaxies (e.g., \citealt{Bartelmann2001PhR}); here, by contrast, we use downsampled simulation particles to directly compute the dark-matter lensing profile around clusters. Each downsampled particle with known position and type is assigned a mass of $10^9~M_\odot$.

        We adopt a pre-computation-and-stacking approach to measure and compare halo lensing profiles across subsamples. For each massive halo, we used \texttt{halotools} (\citealt{Hearin2017AJhalotools}) to compute $\Delta\Sigma(R)$ from downsampled particles ($m_{\rm p}\ge 10^{9}~M_\odot$) within a cube of half-width $40~{\rm Mpc}/h$ centered on the FoF group. Particles are projected along the simulation {\tt XY} direction and binned into 17 logarithmically spaced radial bins out to $17~{\rm Mpc}/h$, from which we computed the mean surface mass density and hence $\Delta\Sigma(R)$. We repeat the same calculation for the dark matter, gas, and stellar components. Because individual-halo profiles, especially at large radii, exhibit substantial scatter due to the low downsampling rate, we stack halos within each subsample by averaging $\Delta\Sigma$ in each radial bin and estimate uncertainties based on 5000 bootstrap resamples.

    \subsection{Top-$N$ Test and difference-set test}
        \label{ssec:test}

        In observations, it is not yet feasible to assemble a halo-mass-complete sample, trace its assembly history, and precisely measure secondary halo properties (e.g., \citealt{Allen2011ARA&A}) as we can in simulations. Instead, cluster samples are selected by observables such as richness and central-galaxy stellar mass, which are incomplete at the low-halo-mass end. To connect our simulation results with such observational selections, we employ two complementary tests:

        \begin{itemize}
            \item \textbf{Top-$N$ test} (\citealt{Huang2022}; as in \citetalias{Xu2024paper1}): we rank halos by a halo-mass proxy and keep the top $N$, mirroring observational cluster selection. The mean and scatter of the resulting halo-mass distribution measure the proxy's quality; a higher mean and smaller scatter at fixed $N$ indicate a better proxy. Observationally, we infer the mean mass and scatter from the stacked lensing profile, whereas in \tng{} we know each halo's mass and compute the scatter directly. We build samples at $N=200,400,600,800$\footnote{For TNG300, these number density thresholds approximately translate into $\log_{10}[M_{\rm halo}/M_\odot]=[14,13.8,13.65,13.55]$.}.

            \item \textbf{Difference-set test}: for two Top-$N$ samples selected by \emph{different} proxies, we take their two difference-sets, the halos belonging to one sample but not the other, and compare their stacked or averaged observables (e.g., lensing profiles) and halo properties (e.g., mass accretion history, concentration). When the two proxies trace halo mass equally well, the difference-sets share nearly the same halo-mass distribution, so any residual difference isolates the imprint of secondary properties and assembly history at fixed halo mass.
        \end{itemize}

        The scatter that makes these tests informative, namely, measurement errors (e.g., \citealt{Mobasher2015ApJ}), projection effects (e.g., \citealt{Wu2022MNRAS}), and, more fundamentally, the diverse assembly histories at fixed halo mass (e.g., \citealt{Bradshaw2020MNRAS}), also means Top-$N$ samples from different proxies never perfectly match (e.g., \citealt{Kugel2024MNRAS}; \citealt{Kwiecien2025PhRvD}), which is exactly what the difference-set test exploits. While both tests apply directly to observations, the observational versions are confounded by measurement errors; we therefore use {\tt TNG300} to control these uncertainties and isolate the \emph{intrinsic} effect, comparing the mass accretion histories and lensing profiles of the difference-sets selected by central stellar mass ($M_\star$-only sample) and richness ($\lambda$-only sample).
    


\section{Results} 
    \label{sec:results}

    In this section, we detail our principal findings. We start in \S\ref{ssec:topn} with the results of the Top-$N$ tests and assess central stellar mass and richness as proxies for \mhalo{}. In \S\ref{ssec:diff}, we present our key results of the difference-set test between the outskirts stellar mass of BCGs and richness. Intrinsic correlations among halo secondary properties, central galaxy stellar mass, and richness are detailed in \S\ref{ssec:rsplane}.

    \subsection{Stellar Mass vs. Richness in the Top-$N$ Test}
        \label{ssec:topn}

\begin{figure}[ht]
        \centering
        \includegraphics[width=1.05\linewidth,trim=2cm 0 0 0]{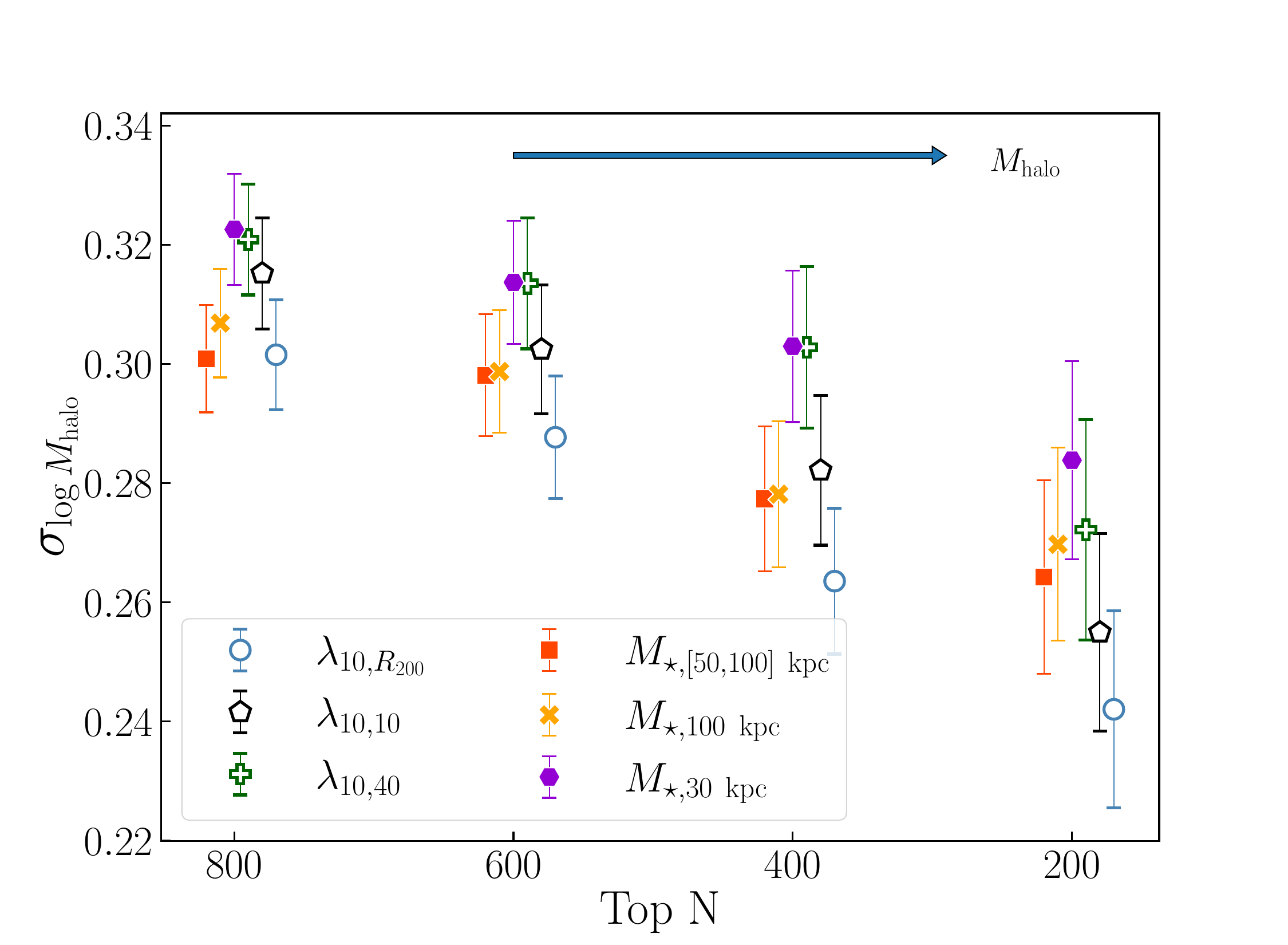}
        \caption{
            The scatter of halo mass distributions in several ``Top-N" bins of {\tt TNG300} halos based on stellar mass or richness as a halo mass proxy. The X-axis shows the sample size $N$, and we shift the points horizontally for better visibility. We use solid symbols to denote stellar mass, whereas open symbols denote richness. Outskirt stellar mass shows a similar level of scatter to richness, given the projection effect. The error is calculated by bootstrapping. 
            The \texttt{Jupyter} notebook for reproducing this figure can be found here: \href{https://github.com/Xuchuyi/cenvssat/blob/main/FigureNotebooks/Fig1acd.ipynb}{\faGithub}.
        }
        \label{fig:topN}
        \end{figure}

    Following the method mentioned in \S\ref{ssec:test}, we apply the Top-$N$ tests on a series of central galaxy stellar masses and richnesses. We present the scatter of the halo mass distributions of these Top-$N$ samples in Fig.~\ref{fig:topN}. This scatter can help us judge whether two massive halo samples have similar halo distributions and evaluate the performance of different halo mass proxies. 

    We adopted 3 definitions of richness in selecting Top-$N$ samples. They are $\lambda_{10,R_{200}}$, $\lambda_{10,10}$, $\lambda_{10,40}$ (see \S\ref{ssec:richmea}). These choices include richness that varies with the projection effect. For the central galaxy stellar mass, we use the mass in a small aperture ($M_{\star,30~{\rm kpc}}$), a large aperture ($M_{\star,100~{\rm kpc}}$), and an outskirt region ($M_{\star,[50,100]~{\rm kpc}}$). For simplicity, these stellar masses are hereafter denoted as $M_{\star,30}$, $M_{\star,100}$ and $M_{\star,[50,100]}$. 

    As in \citetalias{Xu2024paper1}, we evaluate the central galaxy stellar mass and richness performance in tracing halo mass based on the Top-$N$ test scatter. Smaller scatter at fixed $N$ or fixed $N$-bins improves performance. For the central galaxy stellar mass, stellar mass within a small 30 kpc aperture consistently performs worse as a halo mass proxy compared to large-aperture (100 kpc) or outskirts ([50,100] kpc) over all bins in Fig.~\ref{fig:topN}, consistent with previous results (e.g., \citealt{Huang2022}). By contrast, for richness, we find that a stronger projection effect (i.e., counting satellites within a longer cylinder) degrades its performance as a halo mass proxy.

    When comparing the performance of richness with central galaxy stellar mass, we find that the halos selected based on outskirt stellar mass ($M_{\star,[50,100]}$) or large aperture stellar mass ($M_{\star,100}$) generally show comparable scatter with richness-selected halos. Specifically, these two stellar-mass-selected halo samples have similar, or even smaller, mass scatter compared to halo samples selected by richness affected by projection effects ($\lambda_{10,10}$ and $\lambda_{10,40}$). \change{More importantly, when Top-$N$ samples including halos with group-like masses (such as $N=600,~800$ in Fig. \ref{fig:topN}), outskirt stellar mass selections yield scatter comparable to real richness ($\lambda_{10, R_{200}}$). This suggests that the outskirt stellar mass of the central galaxy performs similarly to richness as a halo mass proxy in the group-scale regime, consistent with the findings in \citet[][see their Fig. 8]{Huang2022}.}

    The similar scatter in halo properties between halos selected by outskirts stellar mass and richness suggests that the halo mass distributions of these samples are comparable\change{(\citealt[see their Sec 2.2]{Huang2022})}. As a result, a difference-set test on these sample pairs yields two samples with similar halo mass distributions within a narrow mass range. This provides a useful framework for studying how halo assembly history affects galaxy properties.

    \subsection{Difference between Richness- and Outskirt-Stellar-Mass-Selected Halos}
        \label{ssec:diff}

    Based on the scatter presented in \S\ref{ssec:topn}, we now apply the difference-set test on several pairs of central galaxy stellar mass and satellite richness. In this section, we focus primarily on the results between outskirts stellar mass ($M_{\star,[50,100]}$) and richness, with or without projection effect ($\lambda_{10,40}$, $\lambda_{10,10}$, $\lambda_{10, R_{200}}$, see Tab. \ref{tab:richness}) while the Top-$N$ sample size is $600$. In this case, the scatter associated with halos selected by outskirts stellar mass and richness is comparable across all three richness measures (as in Fig. \ref{fig:topN}). The results of the difference-set tests for other combinations of central-galaxy stellar mass, richness, and sample size are discussed in \S\ref{ssec:masschoice} and \S\ref{app:diffset}.

    As noted above, this comparable scatter implies similar halo-mass distributions, allowing the difference-set test to isolate the aspect in which the two samples genuinely differ: their halo assembly histories. Figure~\ref{fig:figkey} presents the key results of the difference-set test between the outskirt-stellar-mass sample ($M_\star$-only) and the projection-free richness sample ($\lambda$-only); we adopt the projection-free richness ($\lambda_{10, R_{200}}$) here because it most cleanly reveals the intrinsic correlation among central galaxies, satellite galaxies, and halo assembly history. Panel (a) shows how the two difference-sets (red and blue points) populate the stellar mass--richness plane, with their sizes given in the Venn diagram: for the Top-$N=600$ selection, 210 of the 600 halos fall in each difference-set, a smaller fraction than in previous observational studies (e.g., \citealt{Kwiecien2025PhRvD}), possibly reflecting additional systematics in the observations. The inset of panel (c) compares the halo-mass distributions: although the $\lambda$-only sample has a slightly higher mean halo mass, the bulk of both samples overlap. The key point is that, at comparable halo mass, the two samples can nonetheless exhibit markedly different stellar mass distributions, as the example maps in panel (b) illustrate: the $\lambda$-only halos host numerous satellites and would typically be identified as galaxy clusters in observations, whereas the $M_\star$-only halos, despite their similar halo masses, contain fewer satellites but a more prominent extended stellar halo around the central galaxy, and would more likely be classified as galaxy groups.

    To quantitatively assess these structural differences, stacked lensing profiles provide direct insights into the matter distribution and secondary properties of massive halos (e.g., \citealt{Xhakaj2022MNRAS}). As shown in Fig. \ref{fig:figkey}c, the $M_\star$-only and $\lambda$-only samples exhibit distinct profile shapes despite having similar halo masses: the $M_\star$-only sample shows a stronger signal in the inner region ($\lesssim 0.3~{\rm Mpc/h}$), while the $\lambda$-only sample displays a more pronounced peak in the 1-halo regime ($\sim 0.5~{\rm Mpc/h}$). These differences in the 1-halo regime suggest variations in secondary halo properties at fixed halo mass, such as concentration ($c_{200c}$), accretion rate ($\Gamma_{\rm dyn}$) and the offset between the halo center and its center of mass ($x_{\rm off}$): $M_\star$-only halos have higher concentrations but lower accretion rates on average compared to $\lambda$-only halos, which contributes to a stronger central lensing signal (see \S\ref{ssec:off} for further discussion on the offset $x_{\rm off}$).

    \begin{figure*}[htb]
    \centering
    \includegraphics[width=1.05\linewidth,trim=4cm 0 0 0]{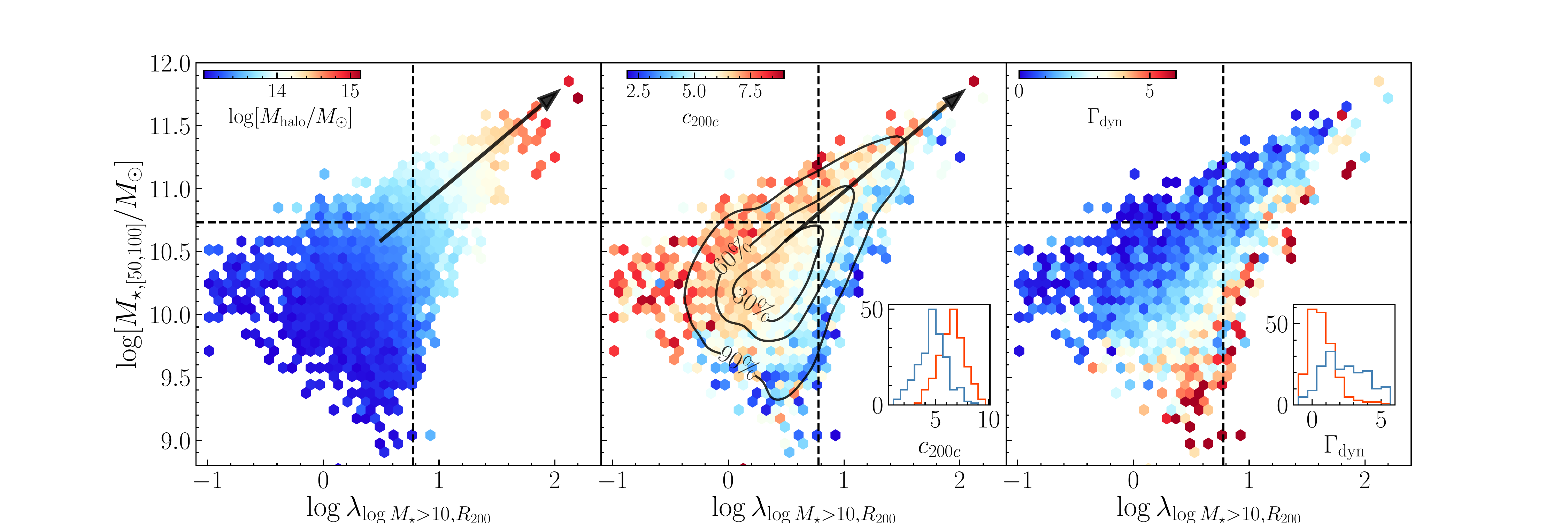}
    \caption{
        The distribution of massive halos' properties on the plane of richness and the outskirts' stellar mass of the central galaxy. Different subfigures are color-coded by different halo properties: halo mass ({\bf Left}), concentration ({\bf Middle}), accretion rate ({\bf Right}).
        These panels reveal a systematic trend in halo concentration and accretion rate on this plane, orthogonal to the trend in halo mass (the black arrow).
        The vertical and horizontal dashed lines show the difference-set test (Fig.\ref{fig:figkey}) selection cut on the richness and stellar mass, and the subfigures in the panels except the first one show the different concentration and accretion rate of the two samples from this test. 
        The \texttt{Jupyter} notebook for reproducing this figure can be found here: \href{https://github.com/Xuchuyi/cenvssat/blob/main/FigureNotebooks/Fig3.ipynb}{\faGithub}.
        }
    \label{fig:planetrend}
    \end{figure*}

    These secondary-property differences stem from the distinct assembly histories of the two samples, establishing a causal chain from halo assembly through secondary properties to the observable lensing signal. As Fig.~\ref{fig:figkey}d shows, the $M_\star$-only halos form earlier (at $z\sim1.2$) than the $\lambda$-only halos (at $z\sim0.8$): their mass accretion histories diverge most strongly around $z\sim1$, with the gap narrowing toward lower redshift and falling within the $1\sigma$ scatter of each sample by $z\sim2$. Through the well-established concentration--formation-time correlation (e.g., \citealt{Wechsler2002ApJconc}), this earlier formation directly sets the higher concentration of the $M_\star$-only halos, while their lower accretion rate reflects comparatively slow late-time growth. These two secondary properties are precisely what the stacked lensing profile encodes: the higher concentration produces the stronger inner (1-halo) signal, and the differing accretion rates imply different splashback radii that should imprint on the transition regime of the profile (e.g., \citealt{Diemer2017ApJRspDep}), a prediction testable with higher signal-to-noise lensing measurements (e.g., \citealt{Diemer2023MNRASRsp}).

    In the outer region beyond $\sim2~{\rm Mpc}/h$ (the 2-halo regime), the stacked profile of the $\lambda$-only halos is instead stronger than that of the $M_\star$-only halos. Although this difference is weaker than in the inner region, it indicates a difference in the large-scale environment between the two samples. Given their statistically distinct assembly histories, this $2\sigma$ difference in the outer $\Delta\Sigma$ profile may hint at assembly bias (e.g., \citealt{Zu2022MNRAS}), though larger simulations and observations are needed to confirm its robustness.

    In summary, the difference-set test exposes a coherent physical sequence: at fixed halo mass, the earlier-assembling $M_\star$-only halos are more concentrated and accrete more slowly today, and these secondary properties are imprinted directly on their stacked lensing profiles. Early assembly also shapes the stellar distribution. An earlier-forming halo has had longer time to disrupt the satellites it accreted, so it retains fewer surviving satellites but builds a more prominent central stellar halo, as we detail in \S\ref{ssec:scenario}. Because the relevant lensing profiles are directly measurable, the difference-set test thus provides a practical observational tool for linking the stellar content of massive halos to their assembly histories at fixed halo mass.

    \subsection{Halo Properties across the Stellar Mass - Richness Plane}
        \label{ssec:rsplane}

    The results of the difference-set test indicate that halo secondary properties, such as halo concentration and accretion rate, are intrinsically correlated with the stellar properties of galaxy clusters. \change{This correlation persists even after controlling for halo mass.} This motivates us to explore how different halo properties are distributed on the outskirts stellar mass ($M_{\star,[50,100]}$) - real richness ($\lambda_{10, R_{200}}$) plane, as illustrated in Fig. \ref{fig:planetrend}.

    We present the distribution of halo mass ($\log[M_{\rm halo}/M_\odot]$), concentration ($c_{200c}$), and accretion rate ($\Gamma_{\rm dyn}$) on the stellar mass–richness plane in the figure. While halo mass increases with richness and stellar mass, concentration and accretion rate show clearly different trends. This suggests an intrinsic correlation between the stellar properties of galaxy clusters and secondary properties of massive halos: at fixed mass, higher-concentration halos tend to host fewer satellites but more prominent stellar halos.
    
    \begin{figure*}[htb]
    \centering
    \includegraphics[width=0.9\linewidth,trim=2cm 0 0 0]{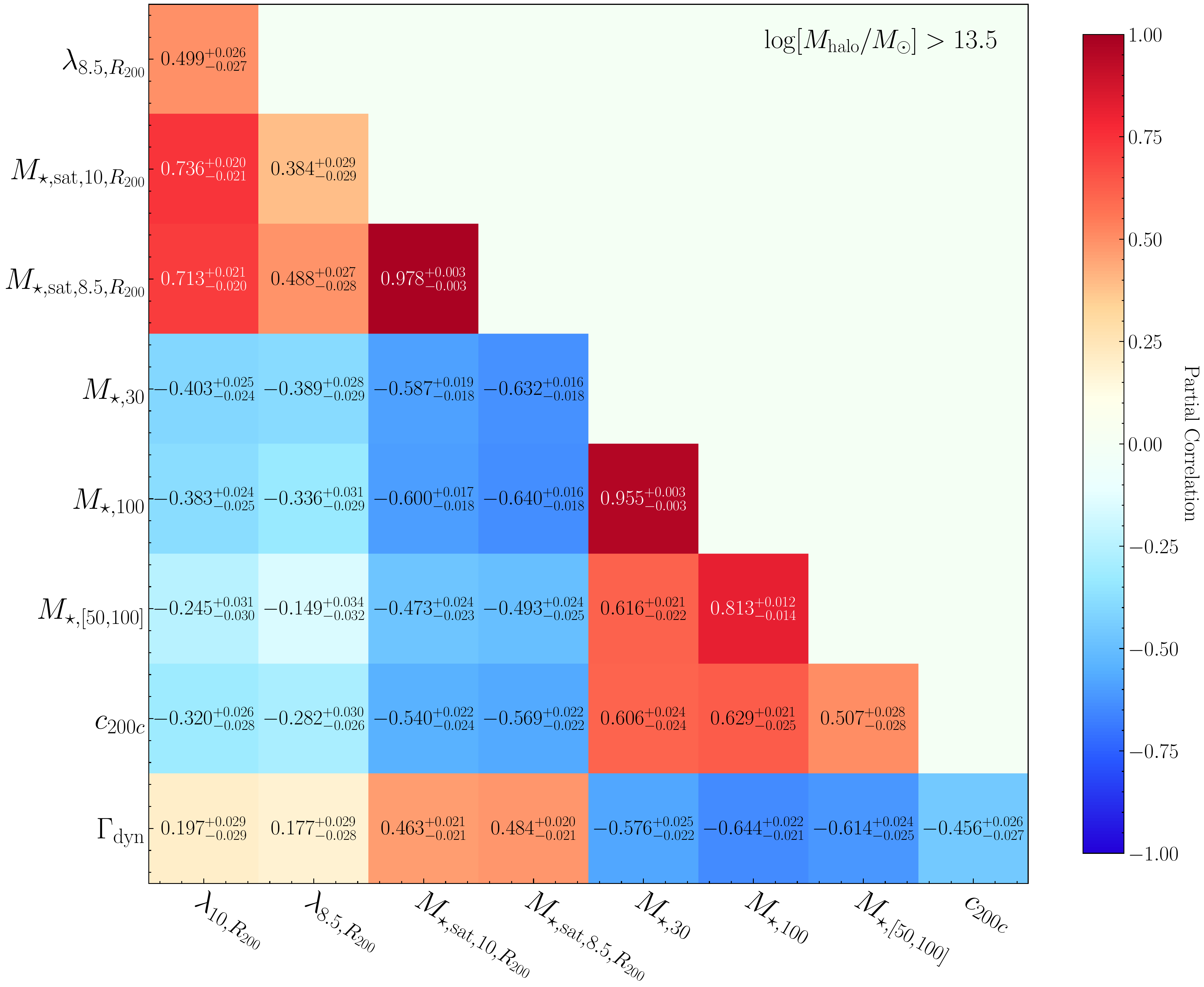}
    \caption{
        Partial correlation matrix of halo properties (richness, total satellite stellar mass, central stellar mass, halo concentration, and accretion rate) at fixed halo mass for massive halos in {\tt TNG300} with $\log[M_{\rm halo}/M_\odot]>13.5$. Each grid shows the Pearson partial correlation coefficient and its standard error between the X-axis and Y-axis properties, with halo mass as the control variable. The error is calculated by bootstrapping. The \texttt{Jupyter} notebook for reproducing this figure is available here: \href{https://github.com/Xuchuyi/cenvssat/blob/main/FigureNotebooks/Fig4.ipynb}{\faGithub}.
        }
    \label{fig:partialprop}
    \end{figure*}

    \change{To be more specific, we calculate the partial correlation between these properties for all halos with mass larger than $10^{13.5}~M_\odot$, using $\log[M_{\rm halo}/M_\odot]$ as the control, as shown in the correlation matrix in Fig. \ref{fig:partialprop}.}\footnote{\change{Here we used Pearson correlation, and when using Spearman correlation, their signs and relative strengths remain unchanged.}} \change{Note that this procedure statistically removes the linear dependence on halo mass, rather than comparing halos within a narrow mass bin as done in the difference set test.}
    For the satellite properties, we consider both the real abundance of satellites and their total stellar mass, as the latter is more closely linked to the host halo mass. \change{We also consider the satellite population with $M_{\star}\geq10^{8.5}$ to better characterize the intrinsic correlation among halo properties, central galaxy stellar mass, and the whole satellite population.}

    \begin{figure*}[htb]
    \centering
    \includegraphics[width=1\linewidth,trim=1cm 0 0 0]{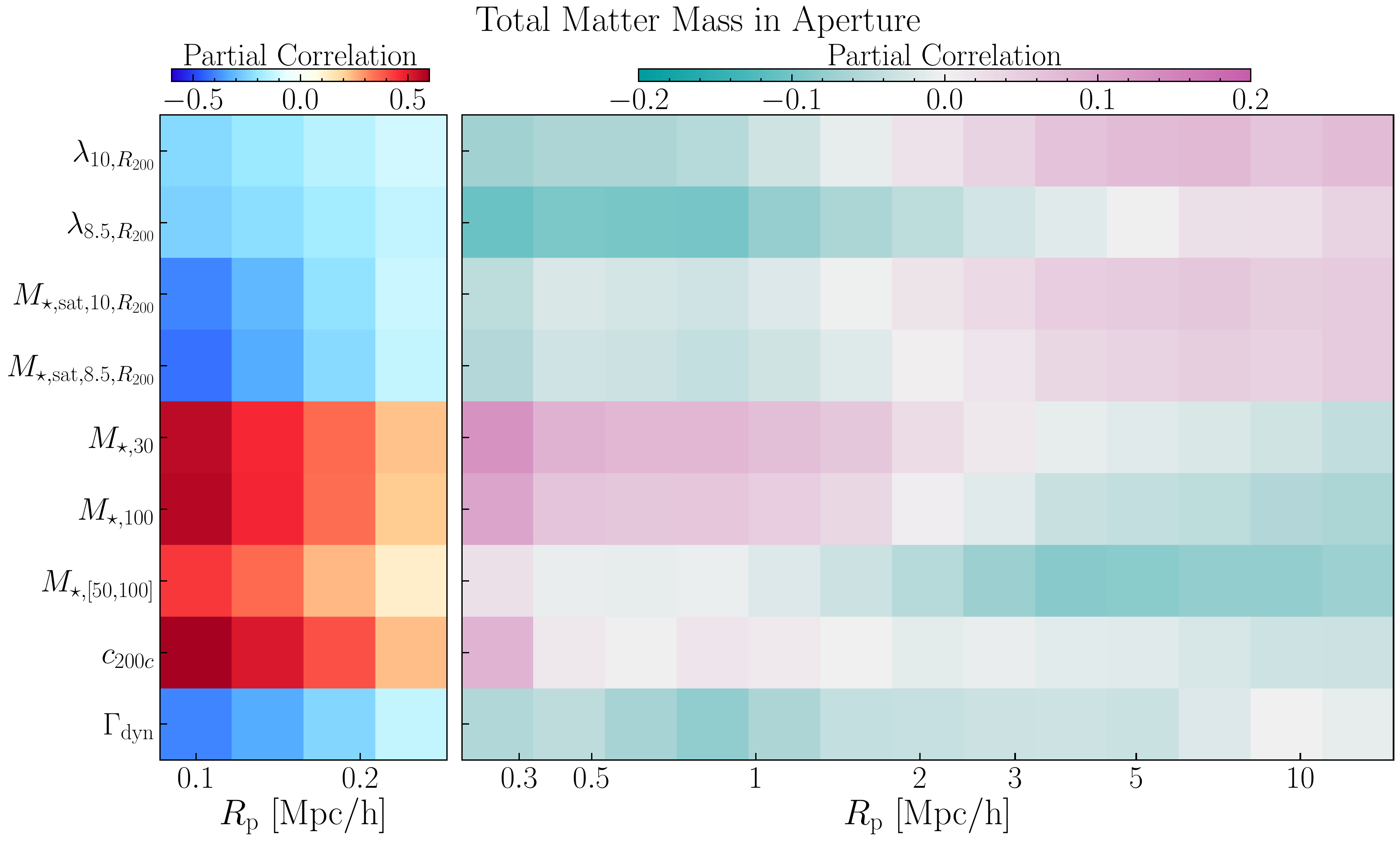}
    \caption{
        Partial correlation matrix between different halo properties, same as Fig.\ref{fig:partialprop}, and total matter mass in different 2D apertures \change{controlling for halo mass}. Each row shows the Pearson partial correlation coefficient between a halo property and total matter mass across 17 apertures with $R_p\in[0.1,17]{~\rm Mpc/h}$, with equal spacing in log-space. \textbf{Left Panel}: For the central region ($<0.25$ Mpc/h), the halo mass in different apertures shows positive partial correlations with the stellar mass of central galaxies and halo concentration, while showing negative correlations with richness, satellite stellar mass, and accretion rate. \textbf{Right Panel}: For the outer region ($R_{\rm p}\in[0.25,17]$ Mpc/h), the partial correlation is weak, but there is a systematic trend from positive correlation to negative correlation for central galaxy stellar mass ($M_{\star,30}$, $M_{\star,100}$, $M_{\star,[50,100]}$).
        The \texttt{Jupyter} notebook for reproducing this figure can be found here: \href{https://github.com/Xuchuyi/cenvssat/blob/main/FigureNotebooks/Fig5.ipynb}{\faGithub}.
        }
    \label{fig:partiallens}
    \end{figure*}

    These partial correlations reveal a coherent physical link between halo and galaxy assembly, which is consistent with Fig. \ref{fig:planetrend}. At fixed halo mass, richness and total satellite stellar mass act as indicators of late-forming halos, which typically exhibit lower $c_{200c}$ and higher $\Gamma_{\rm dyn}$. In contrast, the stellar mass in the central galaxy ($M_{\star,{\rm cen}}$) traces the early-time halo mass growth, showing a negative correlation with $\Gamma_{\rm dyn}$ and a positive correlation with $c_{200c}$. The anti-correlation between satellites and centrals suggests that the two are primary contributors to the present halo mass. We discuss this scenario further in \S\ref{ssec:scenario}.

    More interestingly, the spatial distribution of the central galaxy stellar mass provides further insights: while the inner stellar mass ($M_{\star,30}$) is more strongly correlated with the halo concentration, the outer stellar mass ($M_{\star,[50,100]}$) shows a tighter connection to the recent accretion rate. This indicates the different connection between inner/outer stellar mass and halo assembly history, and we will discuss more about this in \S\ref{ssec:masschoice}. 
    
    In addition, the lensing profiles from the difference-set tests tell us that different halos on the central stellar mass-richness plane have different halo mass profiles. Therefore, we also calculate the partial correlation between the above properties and halo mass in different 2D apertures at fixed halo mass, as shown in Fig. \ref{fig:partiallens}. We find that correlations in the central region ($R_{\rm p} \lesssim 0.25~{\rm Mpc}/h$) are generally stronger than those in the outer region. In the central region, the stellar mass and concentration of the central galaxy are positively correlated with halo mass, whereas richness, total satellite stellar mass, and recent mass accretion rate are negatively correlated. In the outer region, although correlations are weaker, their signs show a clear systematic transition. Specifically, the correlation of central-galaxy stellar mass changes shifts from positive in the inner region to negative in the outer region, whereas satellite-related quantities transition from negative to positive correlations. However, the radius at which this transition occurs differs between Pearson’s and Spearman’s correlation coefficients, highlighting sensitivity to the choice of statistical measure.
    


\section{Discussion} 
    \label{sec:discussions}
    
\subsection{What causes the Differences on the Central Stellar Mass - Richness Plane?}
\label{ssec:scenario}

\begin{figure}[tb]
    \centering
    \includegraphics[width=0.9\linewidth,trim=1.5cm 0 0 0]{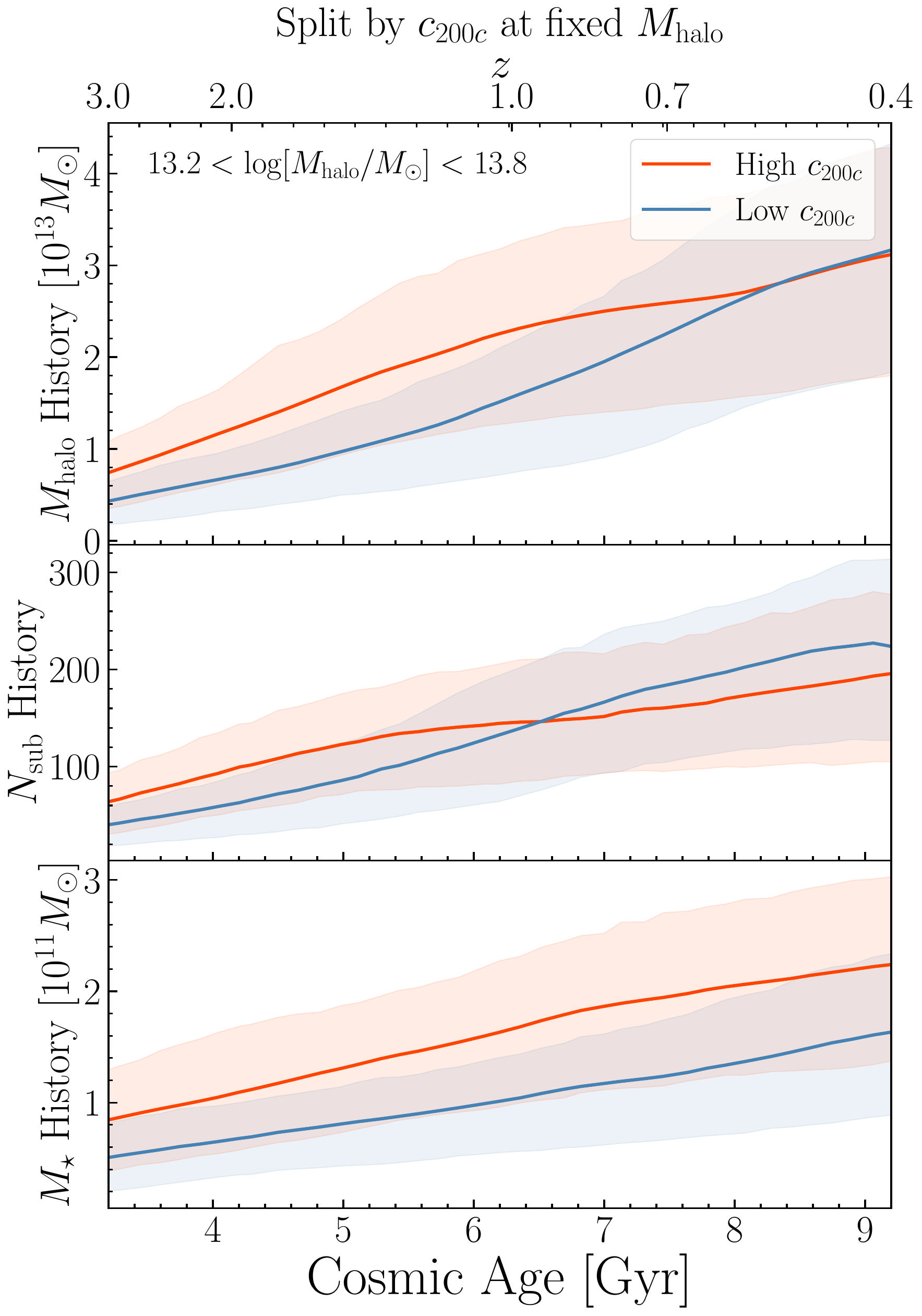}
    \caption{
        Illustration of how massive halos with different concentrations evolved and were located at different loci on the central stellar mass - satellite richness plane.
        {\bf Upper Panel:} Halo mass history ($M_{\rm halo}$).
        {\bf Middle Panel:} Evolution of the number of subhalos, used as a tracer for richness history.
        {\bf Bottom Panel:} Evolution of total stellar mass in the central subhalo, representing the growth of the central galaxy's stellar mass.
        The \texttt{Jupyter} notebook for reproducing this figure can be found here: \href{https://github.com/Xuchuyi/cenvssat/blob/main/FigureNotebooks/Fig6.ipynb}{\faGithub}.
        }
    \label{fig:scenario}
\end{figure}

    Our results show that, at fixed halo mass, halos at different loci on the central stellar mass--richness plane have distinct assembly histories: the $M_\star$-only halos formed earlier and are more concentrated, while the $\lambda$-only halos formed later and host more satellites. Assuming these differences are not driven by baryonic effects (\S\ref{ssec:baryon}), we now ask \emph{how a halo's assembly history shapes the stellar distribution within it}.

    \change{In the ``two-phase'' picture \citep{Oser2010ApJ}, a massive central galaxy and its extended stellar halo grow at $z\lesssim2$ mainly through the accretion of satellite galaxies, which are themselves delivered as the halo accretes smaller halos \citep{Zhao2003MNRAS}. Here, accretion refers to the growth of the whole central stellar component, i.e., the central galaxy plus its extended stellar halo, and it proceeds through two channels: mergers, which deposit satellite stars directly into the central galaxy, and tidal stripping and disruption of orbiting satellites, which preferentially build up the extended stellar halo.}
    Because the total stellar mass of a halo is tightly tied to its halo mass (e.g., \citealt{Bradshaw2020MNRAS}) and is largely accreted (e.g., \citealt{RodriguezGomez2016MNRAS}), two halos of similar mass host a similar \emph{total} stellar mass; what differs is how that mass is \emph{distributed} between surviving satellites and the central galaxy plus its extended stellar halo. This distribution is governed by how much of the accreted satellite population has been disrupted, and the key factor is the time available for disruption: a halo that formed earlier accreted its satellites earlier, leaving more time for dynamical friction and tidal disruption to dissolve them into the central galaxy and its stellar halo (e.g., \citealt{Montenegro-Taborda2025MNRAS}). Since concentration correlates with formation time (e.g., \citealt{Wechsler2002ApJconc}; \citealt{Correa2015MNRAS}; \change{\citealt{Manuwal2025MNRAS}}), the more concentrated $M_\star$-only halos are precisely those that accreted and disrupted their satellites earliest.

    Figure~\ref{fig:scenario} sketches the resulting evolutionary paths for two samples of \change{similar halo mass distribution} but different concentrations, using the number of subhalos to trace richness and the central subhalo's stellar mass to trace the central galaxy. At early times ($z\sim1$--$2$), the earlier-forming halos are already more massive and have accreted more satellites, so their central galaxies grow faster. Toward lower redshift ($z\lesssim1$), the later-forming halos catch up in mass and accrete more satellites, narrowing the mass gap; but because the earlier-forming halos accreted their satellites sooner, those satellites have had more time to be disrupted. The earlier-forming halos thus end with fewer surviving satellites but a more massive central galaxy and a more prominent stellar halo, and are preferentially selected by the $M_\star$ method, whereas the later-forming halos retain more satellites and are preferentially identified as galaxy clusters (the $\lambda$-selected sample).

    This scenario accounts for the correlations between halo secondary properties and stellar distribution seen in Figs.~\ref{fig:planetrend} and \ref{fig:partialprop}, and it links massive galaxies across cosmic time. The difference-set test implies that, if some of the high-redshift massive galaxies or proto-clusters at $z\sim1$--$2$ (e.g., \citealt{Shi2021ApJ}; \citealt{Ito2023ApJ}) are the progenitors of halos that assemble earlier, they are more likely to evolve into a present-day massive galaxy with an extended stellar halo surrounded by a comparatively poor satellite population than into a richer group or cluster of similar halo mass. This connection remains to be tested directly, and we leave a dedicated hydrodynamical-simulation study of these high-redshift progenitors and their low-redshift descendants to future work.

\subsection{Outskirt vs. Aperture Mass}
    \label{ssec:masschoice}

    So far, we have applied the difference-set test using only the outskirt stellar mass $M_{\star,[50,100]}$ (\S\ref{ssec:diff}). To understand how the central galaxy connects to halo assembly history more completely than that single test reveals, we look into the partial-correlation matrix (Fig.~\ref{fig:partialprop}): it shows that the aperture masses $M_{\star,30}$ and $M_{\star,100}$ also correlate with halo concentration, and hence assembly history, at fixed halo mass, even more strongly than $M_{\star,[50,100]}$ in some cases. This motivates extending the difference-set test to these aperture masses and asking how the choice of stellar-mass aperture affects its results.

    To compare these difference-set tests quantitatively, we define quantities measuring the difference between two samples in halo mass, concentration, and lensing profile:
    \begin{enumerate}
        \item $D_{KS}$: the Kolmogorov–Smirnov statistic, the maximum gap between two cumulative distribution functions; here it quantifies the similarity of two halo mass distributions, with $D_{KS}\to0$ for identical distributions.
        \item $\Delta c_{200c}$: the difference between the median halo concentrations of the two samples.
        \item $(\int_{\rm reg}\Delta\Sigma)_{\rm ratio}$: the ratio of the integrated lensing signal $\int_{\rm reg}\Delta\Sigma(R_{\rm p}){\rm d}R_{\rm p}$ between the two samples, evaluated over an inner region $R_p\in[0.1,0.5]~{\rm Mpc/h}$ ($(\int_{\rm inn}\Delta\Sigma)_{\rm ratio}$) and an outer region $R_p\in[2,17]~{\rm Mpc/h}$ ($(\int_{\rm out}\Delta\Sigma)_{\rm ratio}$).
    \end{enumerate}

\begin{figure}[tb]
    \centering
    \includegraphics[width=1\linewidth,trim=2cm 0 0 0]{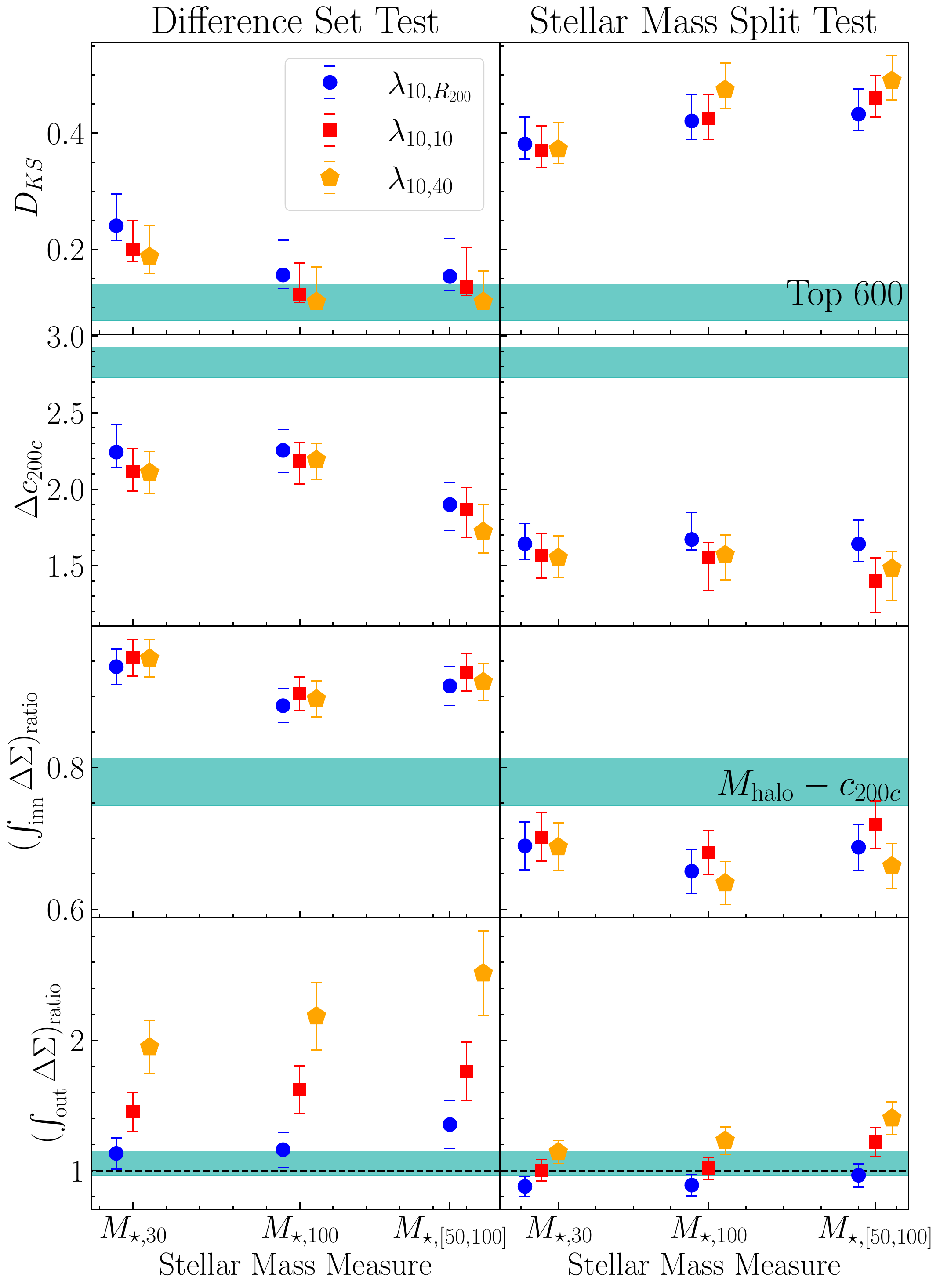}
    \caption{
        Comparison of the performance of different set tests and stellar mass split tests in studying halo assembly history at a fixed halo mass, considering various choices for central galaxy stellar mass ($M_{\star,30},~M_{\star,100},~M_{\star,[50,100]}$) and satellite richness ($\lambda_{10, R_{200}},~\lambda_{10,10},~\lambda_{10,40}$) at Top-$N$ sample size $N=600$. 
        The cyan band is the result of a test by splitting the $M_{\rm halo}-c_{200c}$ plane.
        Four quantities ($D_{KS},~\Delta c_{200c},~(\int_{\rm inn}\Delta\Sigma)_{\rm ratio},~(\int_{\rm out}\Delta\Sigma)_{\rm ratio}$) defined in \S\ref{ssec:masschoice} are used for this comparison. 
        The upper panel ($D_{KS}$) highlights the performance of the different tests in selecting two massive halo samples with similar halo mass, while the lower panels ($\Delta c_{200c},~(\int_{\rm inn}\Delta\Sigma)_{\rm ratio},~(\int_{\rm out}\Delta\Sigma)_{\rm ratio}$) imply the distinct correlations between halo concentration/halo lensing profile and different central galaxy stellar mass/satellite richness, as well as how these respond to the different split tests on the plane. 
        The \texttt{Jupyter} notebook for reproducing this figure can be found here: \href{https://github.com/Xuchuyi/cenvssat/blob/main/FigureNotebooks/Fig7.ipynb}{\faGithub}.
        }
    \label{fig:assess}
\end{figure}

    Figure~\ref{fig:assess} (left column) shows these quantities for difference-set tests across various central stellar masses and richnesses at $N=600$; other sample sizes appear in Appendix \ref{app:diffset}.

    For the halo-mass match ($D_{KS}$), $M_{\star,30}$ shows a larger $D_{KS}$ than $M_{\star,100}$ and $M_{\star,[50,100]}$, so the $M_{\star,30}$–richness difference-set test does not yield mass-matched samples at this number density. The $D_{KS}$ also decreases for richness with stronger projection, consistent with the Top-$N$ test (Fig.~\ref{fig:topN}). The small $D_{KS}$ for $M_{\star,100}$ and $M_{\star,[50,100]}$ confirms that the difference-set test selects halo-mass-matched samples with these proxies in this range\footnote{$N=600$ in TNG300 at $z=0.4$ corresponds to a halo number density of $n \sim 6 \times 10^{-5}~{\rm Mpc}^{-3}$ and a mean halo mass of $M_{\rm halo} \sim 10^{13.6} M_\odot$}.

    This breaks down at higher halo mass: because the stellar-mass-- and richness--halo-mass relations evolve differently (Fig.~\ref{fig:topN}), at $\log[M_{\rm halo}/M_\odot]\gtrsim14$ the difference-set test selects samples with different mass distributions, consistent with the larger $D_{KS}$ at smaller sample sizes in \S\ref{app:diffset}. 

    For the concentration difference $\Delta c_{200c}$, Fig.~\ref{fig:assess} shows that aperture-mass difference-set tests produce larger concentration differences than the outskirt-mass test, consistent with the stronger partial correlation in Fig.~\ref{fig:partialprop}. Since concentration is traced by the inner lensing profile, we examine the inner-region ratio $(\int_{\rm inn}\Delta\Sigma)_{\rm ratio}=(\int_{\rm inn}\Delta\Sigma)_{\lambda}/(\int_{\rm inn}\Delta\Sigma)_{M_\star}$. However, the small-aperture ($M_{\star,30}$) test shows no larger inner-region difference than the outskirt ($M_{\star,[50,100]}$) test, despite its larger concentration difference. This lensing discrepancy arises from the higher halo mass of the $\lambda$-only sample in the $M_{\star,30}$--$\lambda_{10, R_{200}}$ test, again underscoring the need to control halo mass when studying assembly history.

    In the outer region, beyond the projection effect (\S\ref{ssec:projeff}), the difference-set lensing profiles show weak differences (Fig.~\ref{fig:figkey}) that depend on the mass proxy: Fig.~\ref{fig:assess} shows they are consistently more pronounced for outskirt mass than for aperture mass. The two mass definitions may thus couple differently to the large-scale environment, a signature warranting further investigation with larger samples.

    Overall, these differences reflect distinct correlations among the central galaxy's inner and outer regions, the satellites, and the host halo. The inner and outer regions roughly correspond to the BCG and ICL/IGrL: although their boundary is ill-defined (e.g., \citealt{Marini2022MNRAS}) because of their similar low-redshift growth, the two components form via different pathways and show distinct radial trends (e.g., \citealt{RodriguezGomez2016MNRAS}; \citealt{Montenegro-Taborda2023}), with the \exsitu{} fraction\footnote{The ratio of exsitu stellar mass over total stellar mass} rising outward. In IllustrisTNG, BCGs are merger-dominated while the ICL is mainly stripped from surviving satellites (\citealt{Montenegro-Taborda2025MNRAS}), suggesting that halo assembly history is imprinted on the central stellar distribution (e.g., \citealt{Dacunha2025arXiv}), which we will explore in future work.

\subsection{Comparison with a Stellar-Mass Split at Fixed Richness}
    \label{ssec:testchoice}

    \change{
    An alternative way to explore halo assembly history on the central stellar mass--richness plane is to split a richness-selected sample by central-galaxy stellar mass. Following \citet{Zu2021MNRAS}, this ``stellar-mass split'' is performed using a richness-dependent stellar-mass cut: clusters above and below the mean $M_\star$--$\lambda$ relation form the high- and low-$M_\star$ subsamples, respectively. By construction, the two subsamples therefore have the same richness distribution. However, this does not guarantee the same halo-mass distribution, because residuals in $M_\star$ and $\lambda$ at fixed halo mass may be correlated. In the observational analysis of \citet{Zu2021MNRAS}, stacked weak-lensing modeling showed that the two subsamples had consistent average halo masses but different concentrations.
    }

    \change{
    We apply the same type of split to the richness-selected halos in {\tt TNG300} and compare it with our difference-set test in Fig.~\ref{fig:assess}. Because the halo mass of every simulated system is known, $D_{\rm KS}$ directly tests whether the two subsamples have the same halo-mass distribution. At Top-$N=600$, the stellar-mass split produces a substantially larger $D_{\rm KS}$ than the difference-set test for every stellar-mass and richness definition considered. Thus, in {\tt TNG300}, matching the richness distributions does not produce halo-mass-matched subsamples. By contrast, the difference-set test provides substantially better halo-mass matching when either $M_{\star,100}$ or $M_{\star,[50,100]}$ is paired with richness; the match is poorer when $M_{\star,30}$ is used. This conclusion is insensitive to our richness definition and hence to the adopted level of projection effect, although projection strongly changes the contrast between the outer lensing profiles.
    }

    \change{
    The stellar-mass split in Fig.~\ref{fig:assess} also produces a strong inner-lensing contrast. However, the high-$M_\star$ subsample has both a higher halo mass and a higher concentration in {\tt TNG300}. Its enhanced inner lensing therefore cannot be attributed solely to concentration or assembly history. In this simulation, the stellar-mass split does not solve the halo-mass-matching problem and should not be interpreted as preferable to the difference-set test. It can be used to identify variations across the stellar mass--richness plane only if the residual halo-mass difference is independently constrained and included in the interpretation.
    }

    \change{
    The contrast with \citet{Zu2021MNRAS} does not imply that their observational result is incorrect. Instead, it demonstrates that the success of either selection method depends on the joint relation among halo mass, central stellar mass, and richness. In observations, the halo-mass matching must therefore be verified independently, for example through weak lensing and through comparisons between complementary selection methods. The difference between the observational and {\tt TNG300} results may itself constrain how accurately the simulation reproduces the massive-galaxy--halo connection. Ultimately, rather than relying on a particular split to remove halo-mass differences, we aim to model the trends of halo mass and secondary halo properties over the full observed stellar mass--richness plane, analogous to Fig.~\ref{fig:planetrend}.
    } 

    \begin{figure}
    \centering
    \includegraphics[width=0.9\linewidth,trim=0.5cm 0 0 0]{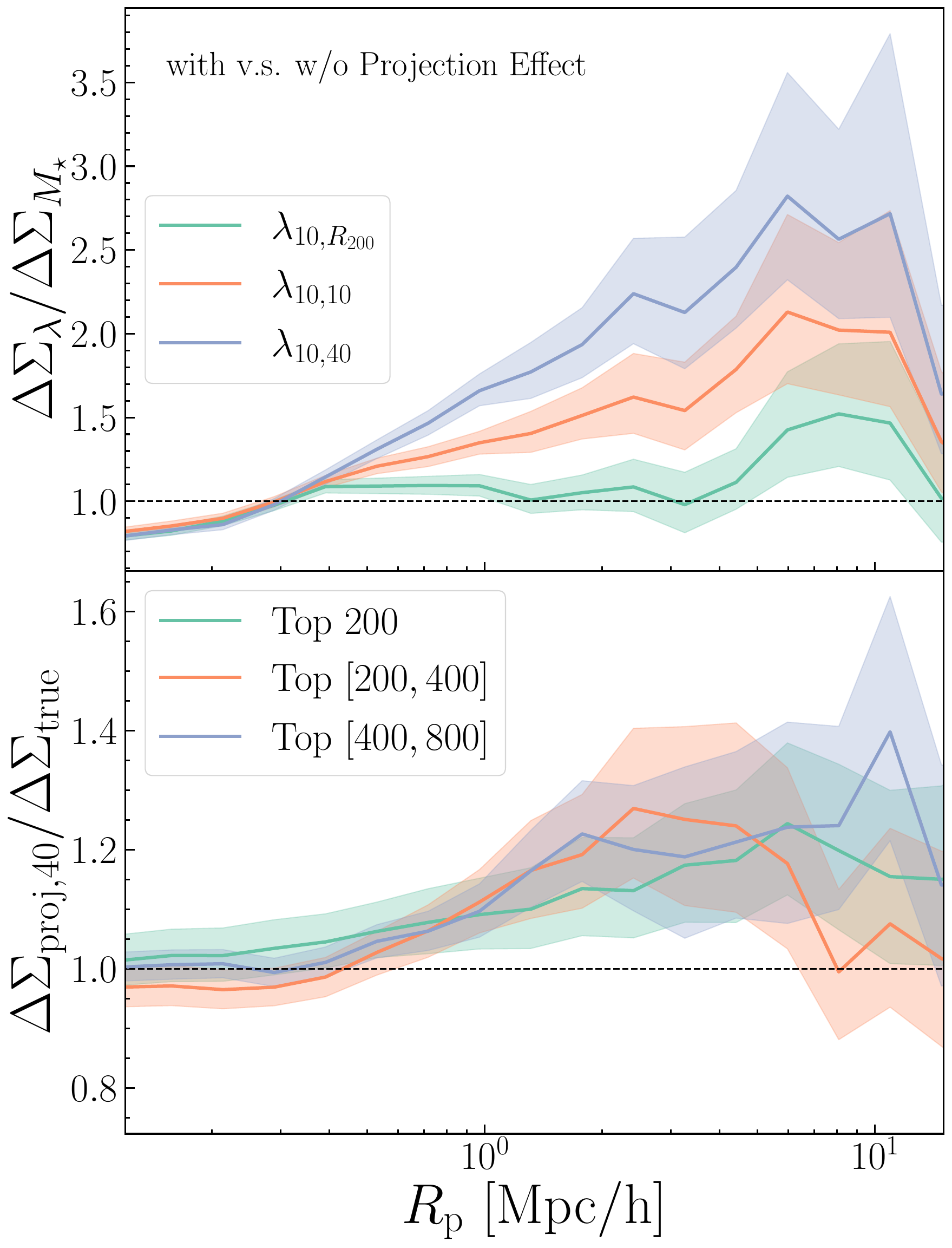}
    \caption{
        Illustration of the projection effect in the lensing profiles.\\ 
        {\bf Upper Panel:} Similar to the bottom part of Fig.\ref{fig:figkey}c, here shows the difference-set test lensing ratio between outskirt stellar mass ($M_{\star,[50,100]}$) and 3 different measures of richness: $\lambda_{10, R_{200}}$ (green), $\lambda_{10,10}$ (orange), $\lambda_{10,40}$ (blue). Compared with richness defined in a sphere, definitions that account for the projection effect (in a cylinder) yield a stronger lensing signal ratio in the outer region.\\
        {\bf Bottom Panel:} The lensing ratio between two samples of halos selected based on different richness measures: one defined in a $\pm40$ Mpc cylinder that considers the projection effect ($\Delta\Sigma_{\rm proj,40}$) and one defined in a sphere that is similar to a true satellite abundance ($\Delta\Sigma_{\rm true}$). The 3 lines correspond to different Top-N selection bins. Compared with a sample selected based on true richness, a sample affected by the projection effect exhibits a stronger lensing signal in the outer region. 
        The \texttt{Jupyter} notebook for reproducing this figure can be found here: \href{https://github.com/Xuchuyi/cenvssat/blob/main/FigureNotebooks/Fig8.ipynb}{\faGithub}.
        }
    \label{fig:projection}
    \end{figure}

\subsection{Projection Effects}
\label{ssec:projeff}

    As noted above, projection effects in observations complicate accurate measurement of true richness (e.g., $\lambda_{10, R_{200}}$). To investigate this, we also test the impact of projection effects on the lensing profile by considering the richness within a cylinder in both the Top-$N$ and difference-set tests.

    In the bottom panel of Fig. \ref{fig:projection}, we compare lensing profiles of halos selected by true richness ($\lambda_{10, R_{200}}$) versus richness influenced by the projection effect ($\lambda_{10,40}$) via their ratio. We find that projection effects enhance the outer region ($R_{\rm p}\gtrsim 1{\rm Mpc/h}$) signal by $\sim10-40\%$ across all number density ranges, consistent with previous studies (e.g., \citealt{Sunayama2020MNRAS}; \citealt{Wu2022MNRAS}). 

    More importantly, we examine how projection effects influence the results of the difference-set test (Fig. \ref{fig:figkey}), as shown in the upper panel of Fig. \ref{fig:projection}. We compare the lensing profile ratios between the difference-sets of halos selected by outskirt stellar mass $M_{\star,[50,100]}$ and richness $\lambda$. Here, we consider the richness under different levels of the projection effect: none ($\lambda_{10, R_{200}}$), weak ($\lambda_{10,10}$), and strong ($\lambda_{10,40}$). We find that as the projection effect becomes stronger, the lensing ratio ($\Delta\Sigma_{\lambda}/\Delta\Sigma_{M_\star}$) increases in the outer region ($R_{\rm p} \gtrsim 0.4~{\rm Mpc/h}$). In the inner region, however, all cases show a similar ratio ($<1$), indicating a higher concentration of $M_\star$-only halos and suggesting that this difference-set test remains a promising tool for studying halo secondary properties and assembly history in observational data, even in the presence of the projection effect. In addition, as Fig. \ref{fig:assess} shows, such results are also consistent when the choice of central stellar mass is $M_{\star,30}$ and $M_{\star,100}$. 

    The projection effect on the lensing profiles reflects the selection bias of massive halos. Since these halos are typically located at the nodes of large-scale filaments, when we use richness measures such as $\lambda_{10,10}$ and $\lambda_{10,40}$ to select massive halos, some halos may not be particularly massive, but their host filaments align along the line of sight (e.g., \citealt{Sunayama2020MNRAS}). Consequently, the enhancement of lensing ratios in the outer region reflects the projected matter distribution of these filaments.

\begin{figure}[t]
    \centering
    \includegraphics[width=0.9\linewidth,trim=1.5cm 0 0 0]{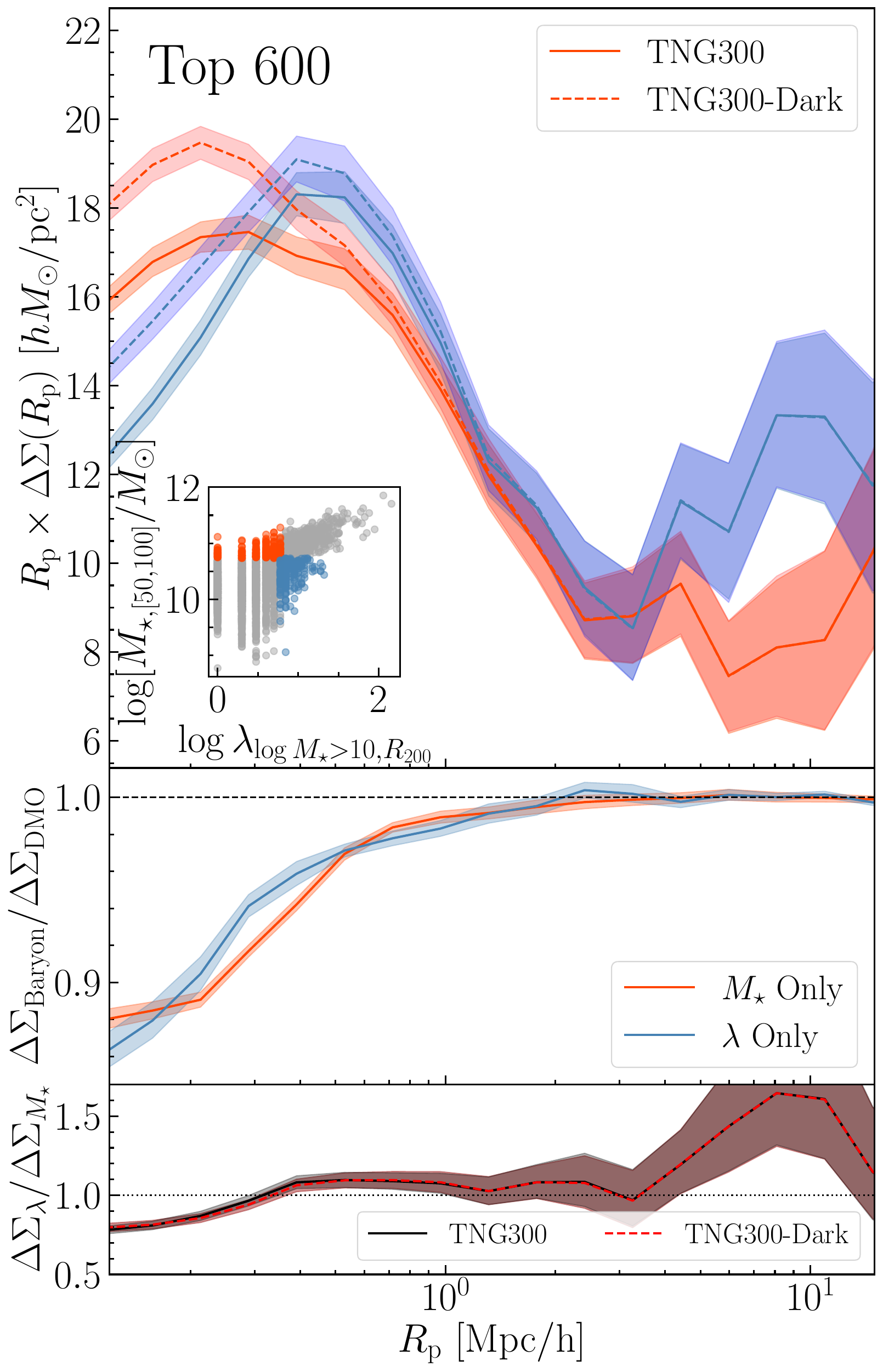}
    \caption{
        {\bf Upper panel:} The lensing profiles of Fig.\ref{fig:figkey} (solid line) and their dark-matter-only version (dashed line) in {\tt TNG300-Dark}.
        {\bf Middle panel:} The ratio of the lensing profile of matched halos in hydro and dark-matter-only simulations.
        {\bf Bottom panel:} Similar to the bottom figure in panel (c) of Fig.\ref{fig:figkey}. The black solid line is for halos in {\tt TNG300}, while the red dashed line is for matched halos in {\tt TNG300-Dark}. They are basically consistent within the bootstrapping error range.
        The \texttt{Jupyter} notebook for reproducing this figure can be found here: \href{https://github.com/Xuchuyi/cenvssat/blob/main/FigureNotebooks/Fig9.ipynb}{\faGithub}.
        }
    \label{fig:bary}
\end{figure}

\subsection{Baryonic Effects}
    \label{ssec:baryon}

    Two primary factors contribute to differences in halo concentration and the lensing profile in the inner region. The first one is the halo assembly history. As discussed in \S\ref{ssec:diff}, halos that form earlier tend to have higher concentrations and stronger lensing signals in the inner region. The second factor is baryon feedback (e.g., \citealt{Mead2010MNRASbaryon}; \citealt{Henson2017MNRASbaryon}; \citealt{Lee2018MNRASbaryon}; \change{\citealt{Beltz-Mohrmann2021ApJ}; \citealt{Shao2023MNRAS}}), which redistributes mass within the halo by expelling gas from the central region, thereby reducing the overall halo concentration. This process results in a shallower inner density profile and a weaker lensing signal. While Fig. \ref{fig:figkey} confirms that the two halo samples have different average halo formation times, it is important to note that baryon feedback may still vary between massive halos, particularly those with different central stellar masses and richness. As a result, baryonic effects could contribute to differences in the lensing profiles in the inner regions.

    To quantify the influence of baryon feedback on the lensing profiles, we compare the lensing profiles of the halo samples in Fig. \ref{fig:figkey} with their dark-matter-only (DMO) counterparts from the {\tt TNG300-Dark} simulation, as shown in Fig. \ref{fig:bary}. The upper panel shows the lensing profiles for the hydro (solid line) and DMO (dashed line) halo versions. In the inner region ($R_p\lesssim0.5$ Mpc/h), both the outskirt-stellar-mass-only and richness-only halos are significantly affected by baryon feedback, leading to a reduction in the lensing signal by approximately 10\%. The middle panel quantifies this suppression through the lensing ratio $\Delta\Sigma_{\rm Baryon}/\Delta\Sigma_{\rm DMO}$, demonstrating stronger baryonic effects in outskirt-stellar-mass-only halos compared to their richness-only counterparts. This differential suppression implies that halos hosting more massive central galaxies undergo stronger baryonic impact. However, although the strength of the baryonic effect differs between these two halo samples, it does not significantly contribute to the difference in their lensing profiles. The bottom panel of Fig. \ref{fig:bary} presents the lensing ratio $\Delta\Sigma_{\lambda}/\Delta\Sigma_{M_\star}$ (as shown in Fig. \ref{fig:figkey}c), and we observe that the ratio remains consistent within the error bar range between the {\tt TNG300} and {\tt TNG300-Dark} simulations.

\subsection{How to select Massive Halos for Cosmology?}
\label{ssec:off}

\begin{figure}[tb]
    \centering
    \includegraphics[width=0.8\linewidth,trim=1.5cm 0 0 0]{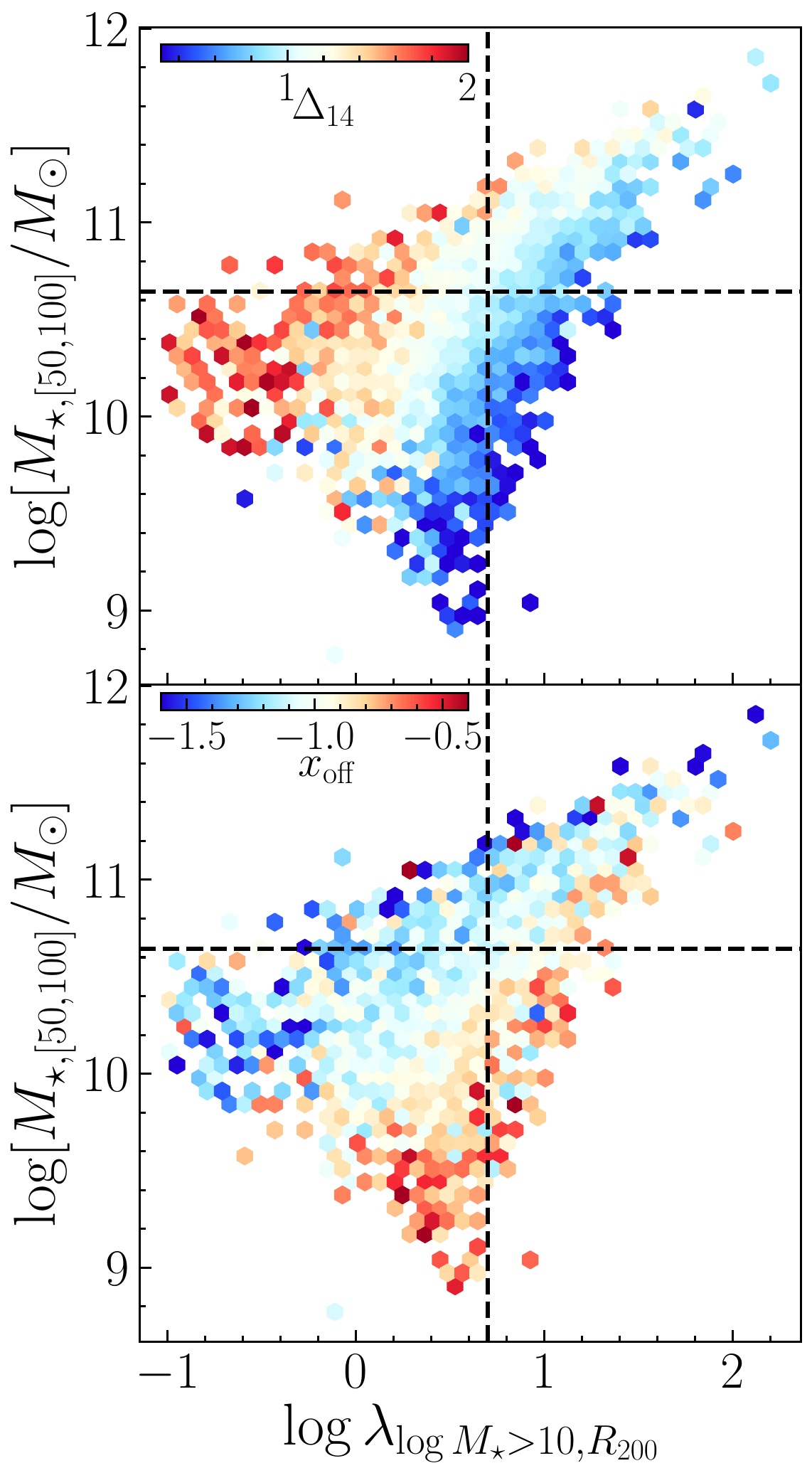}
    \caption{
        Similar to Fig. \ref{fig:planetrend}, but color-coded by different properties.
        \textbf{Top panel}: Color-coded by the stellar mass gap, $\Delta_{14} \equiv \log(M_1/M_\odot) - \log(M_4/M_\odot)$, between the most massive and the fourth most massive galaxies in the cluster.
        \textbf{Bottom panel}: Color-coded by the offset between the minimum of the gravitational potential well and the halo's center of mass, $x_{\rm off}$; here, we use the center of mass of the central subhalo to mitigate influence from the FoF algorithm.
        Both properties reflect the halo's dynamical state and show clear trends on the stellar mass–richness plane. This suggests that richness-based selection favors halos with smaller stellar mass gaps and larger offsets, which are characteristic of unrelaxed systems.
        The \texttt{Jupyter} notebook for reproducing this figure can be found here: \href{https://github.com/Xuchuyi/cenvssat/blob/main/FigureNotebooks/Fig10.ipynb}{\faGithub}.
        }
    \label{fig:obsimp}
\end{figure}

    The difference-set test shows that different selection methods imprint distinct halo secondary properties: richness-selected halos tend to form later, are less concentrated, and accrete faster, whereas outskirts-stellar-mass-selected halos form earlier and are more concentrated. These biases extend to the halo's dynamical state. To demonstrate this, we introduce two further indicators of whether a halo is relaxed: the center-of-mass offset $x_{\rm off}$, the displacement between the halo center\footnote{\change{We adopted \texttt{GroupPos}, the minimum of gravitational potential, as halo center.}} and its center of mass normalized by $R_{200c}$ (as in \citealt{Xhakaj2022MNRAS}), and the stellar mass gap $\Delta_{14}=\log M_{\star,{\rm 1st}}-\log M_{\star,{\rm 4th}}$ between the central and the fourth most massive galaxy, analogous to the observational magnitude gap (e.g., \citealt{Golden-Marx2025MNRAS}); here $M_{\star,{\rm 1st}}=M_{\star,30}$ and satellite masses follow \S\ref{ssec:shcat_trees}. As Fig.~\ref{fig:obsimp} shows, both quantities vary systematically across the stellar mass--richness plane, along the same direction as concentration and accretion rate (cf. Fig.~\ref{fig:planetrend}): richness-based selection preferentially picks halos with smaller mass gaps and larger offsets, i.e., more recently formed and more dynamically unrelaxed systems. Such selection biases can significantly affect cosmological interpretation (e.g., \citealt{Hearin2015MNRASRSDAB}; \citealt{McEwen2018MNRASAB}).

    A key goal for cluster cosmology is to extract information from small-scale ($\lesssim10~{\rm Mpc}$) lensing and clustering (e.g., \citealt{Yuan2022MNRASsmall}; \citealt{Zhai2023ApJsmall}), where modeling is complicated by nonlinearity, baryonic effects, and assembly bias (e.g., \citealt{Hearin2016MNRASdecoratedHOD}; \citealt{Leauthaud2017MNRAS}). Assembly bias, the dependence of clustering and lensing on secondary properties at fixed halo mass (e.g., \citealt{Gao2007MNRAS}; \citealt{Dalal2008ApJ}), is exactly what our results expose: because different selection methods favor different secondary properties, samples of similar halo mass can have different clustering and lensing profiles, so cluster cosmology that ignores this bias may infer systematically biased parameters (e.g., \citealt{Zentner2014}; \citealt{Yuan2020MNRAS}).

    The dynamical-state bias is particularly relevant for weak-lensing cluster cosmology, where halo mass estimates are already biased relative to the truth (e.g., \citealt{Bahe2012MNRAS}; \citealt{Euclid2024A&A}) by triaxiality, projection effects, and miscentring (e.g., \citealt{Clowe2004MNRAS}; \citealt{Gavazzi2005A&A}; \citealt{Simet2017MNRAS}), with the largest biases (up to $\sim60\%$) in unrelaxed clusters (\citealt{Lee2023ApJ}). Since richness-based selection preferentially yields such unrelaxed systems, it can amplify these mass-calibration biases.

    In summary, our {\tt TNG}300 analysis shows that both imaging-based selection methods, satellite richness and central-galaxy outskirt stellar mass, imprint systematic biases on the resulting halo samples, especially in the low-mass, incomplete regime. These biases appear as secondary properties such as formation time, concentration, and dynamical state, and thus propagate into small-scale lensing and clustering signals. Accounting for or mitigating them is therefore important for future cosmological analyses; a promising path is to combine selection methods and model the joint relation between halo properties and the central stellar mass--richness plane to reduce their impact on cosmological inference.



\section{Conclusions and Future Directions} 
\label{sec:summary}

    Recent deep-imaging and weak-lensing observations of low-redshift massive galaxies and clusters hint at an intrinsic connection between the stellar components of massive halos and their assembly histories (e.g., \citealt{Zu2021MNRAS}). To investigate this link with a controlled, mass-complete sample, we use massive halos ($\log[M_{\rm 200c}/M_\odot]\geq13.0$) from the {\tt TNG300} hydrodynamical simulation at $z=0.4$ (\citealt{Pillepich2018MNRAS}), for which the galaxy populations, lensing profiles, peak mass accretion histories, and secondary halo properties are all available. To compare halo properties at fixed halo mass using directly observable quantities, we introduce the \emph{difference-set test}, and we assess its robustness against projection effects and baryonic physics. Our main findings are:

    \begin{itemize}

        \item \textbf{Outskirt stellar mass rivals richness as a halo-mass proxy at group scales.} In the Top-$N$ test (Fig.~\ref{fig:topN}), at $\log[M_{\rm halo}/M_\odot]\sim[13.2,13.8]$ the outskirt ($M_{\star,[50,100]}$) and large-aperture ($M_{\star,100}$) stellar masses show scatter comparable to satellite richness, even the projection-free $\lambda_{10, R_{200}}$. At higher mass ($\sim10^{14}~M_\odot$), central stellar mass underperforms $\lambda_{10, R_{200}}$ but remains competitive with projection-affected richness.

        \item \textbf{At fixed halo mass, the two proxies select halos with different assembly histories.} Applying the difference-set test on the $M_{\star,[50,100]}$--$\lambda_{10, R_{200}}$ plane yields two samples with similar halo-mass distributions (Fig.~\ref{fig:figkey}); their lensing profiles and mass accretion histories reveal that $M_\star$-selected halos are more concentrated and form earlier than richness-selected halos. Optical halo-mass proxies thus imprint a selection bias on halo assembly history.

        \item \textbf{Halo secondary properties vary systematically across the stellar mass--richness plane.} The difference-set test and the partial-correlation analysis (Figs.~\ref{fig:planetrend}, \ref{fig:partialprop}, \ref{fig:partiallens}) expose coherent trends in concentration, accretion rate, and recent mass growth, linking the stellar buildup of central galaxies to the mass-growth history of their host halos (Fig.~\ref{fig:scenario}).

        \item\change{\textbf{Matching richness does not guarantee matching halo mass.} At Top-$N=600$, the stellar-mass split does not produce halo-mass-matched subsamples in {\tt TNG300}, whereas the difference-set test performs substantially better with large-aperture or outskirt stellar mass (Fig.~\ref{fig:assess}). Halo-mass matching must therefore be verified before interpreting assembly-history differences.
        }

        \item \textbf{Projection effects bias only the outer lensing signal.} Projection in the richness measurement enhances the outer-region lensing of richness-selected halos but leaves the inner profile, and hence the concentration contrast, essentially unchanged (Fig.~\ref{fig:projection}).

        \item \textbf{Baryonic physics does not drive the difference-set signal.} Although baryon feedback suppresses the inner lensing profile, the $\Delta\Sigma_{\lambda}/\Delta\Sigma_{M_\star}$ ratio is barely affected in {\tt TNG300} (Fig.~\ref{fig:bary}); baryonic effects nonetheless remain a challenge for modeling the plane with observational data.

    \end{itemize}

    The central stellar mass--satellite richness plane is a promising observational tool for probing halo secondary properties and assembly history. With upcoming deep imaging (e.g., HSC, Euclid, and the Vera C. Rubin Observatory) and wide-field spectroscopy (e.g., DESI), we will measure the diffuse stellar halos and satellite populations of massive halos far more accurately. Building on this, we plan to model the joint relation among the central stellar mass-- satellite richness plane, halo mass, and secondary properties (e.g., concentration, splashback radius), using large samples of massive galaxies and clusters together with higher-quality weak-lensing measurements and larger-volume simulations, and to extend the difference-set test to a broader range of halo masses and observables.

    Our results also highlight a caveat for cosmology: because both the outskirts stellar mass and satellite richness are halo-mass proxies biased toward optical cluster finding, combining them may yield a more accurate and less biased proxy for small-scale cluster cosmology (Xu et al., in preparation). Finally, motivated by the finding that central-galaxy stellar profiles encode information about halo assembly history, we will explore the connection between the full stellar profile of central galaxies and their host-halo assembly history in future work.



\section*{Acknowledgments}

  SX and SH acknowledge the support from the Ministry of Science and Technology of the People’s Republic of China (MOST) Grant No. 2023YFA1605600, the National Natural Science Foundation of China (NSFC) Grant No. 12273015, and NSFC Grant No. 12433003. SH also acknowledges the generous support from Ma Huateng Foundation.

  AL and BD acknowledge support from the Kavli Institute for Theoretical Physics. The National Science Foundation under Grant No. NSF PHY11-25915 and Grant No. NSF PHY17-48958

  BD and KL were partially supported by the National Science Foundation under Grant. No. AST-2206695.

  KL was partially supported by the National Science Foundation Graduate Research Fellowship Program under Grant No. DGE 2236417
  
\software{
    \href{http://www.numpy.org}{\code{NumPy}} \citep{Numpy},
    \href{https://www.astropy.org/}{\code{Astropy}} \citep{astropy},  
    \href{https://www.scipy.org}{\code{SciPy}} \citep{2020SciPy}, 
    \href{https://matplotlib.org}{\code{Matplotlib}} \citep{matplotlib} 
}

\appendix
\twocolumngrid
\renewcommand{\thefigure}{\Alph{section}\arabic{figure}}
\section{Difference-set test at other halo mass ranges}
\label{app:diffset}

\setcounter{figure}{0}

\begin{figure*}[h]
    \centering
    \includegraphics[width=0.9\linewidth,trim=1.5cm 0 0 0]{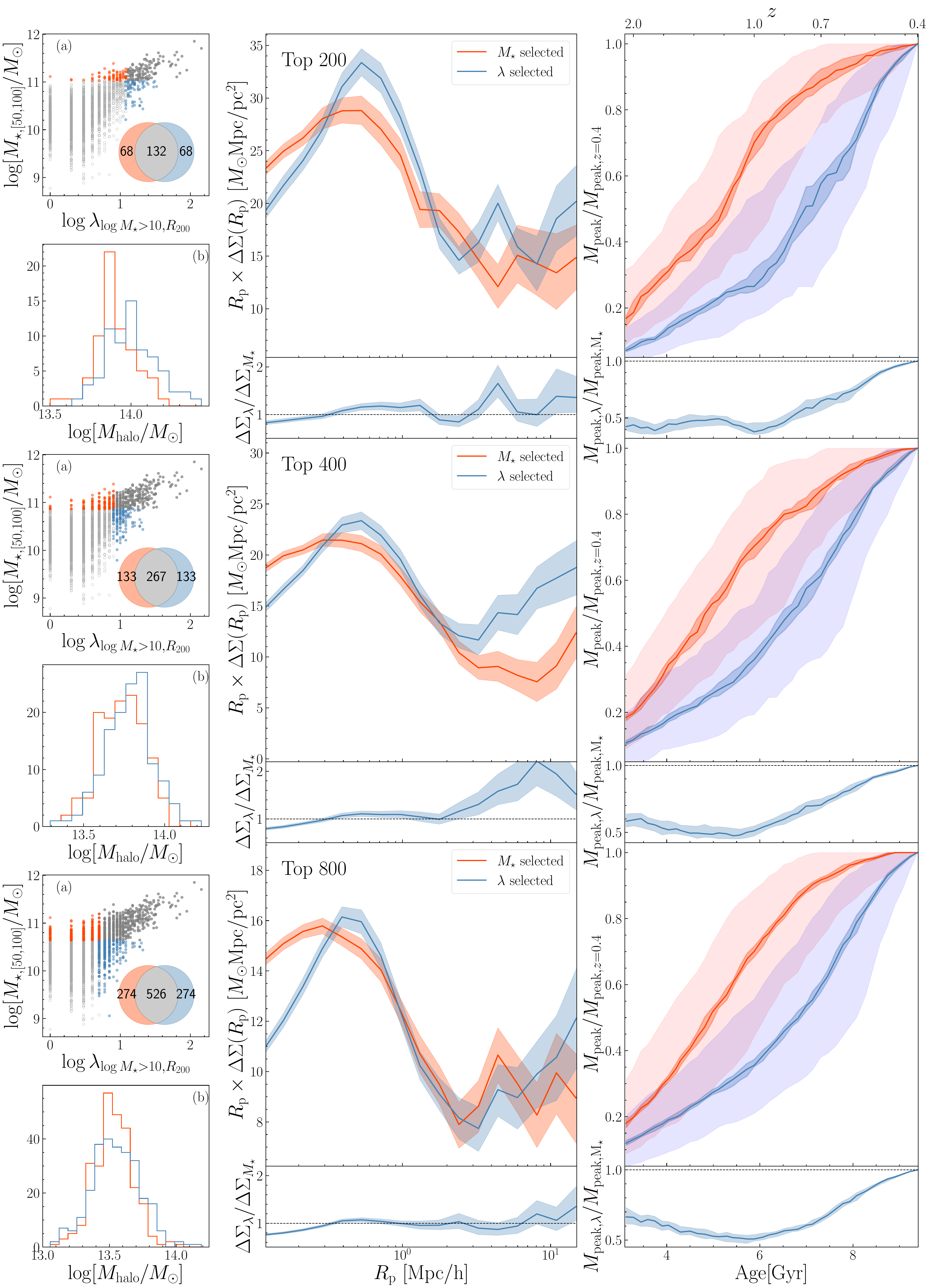}
    \caption{
        Similar to Fig. \ref{fig:figkey}. The difference-set test between outskirt stellar mass ($M_{\star,[50,100]}$) and satellite richness ($\lambda_{10,R_{200}}$) with Top-N sample size $N=200,400,800$. These additional cases illustrate how the test behaves with varying halo mass and provide a more comprehensive view of its performance.
        The \texttt{Jupyter} notebook for reproducing this figure can be found here: \href{https://github.com/Xuchuyi/cenvssat/blob/main/FigureNotebooks/FigA1.ipynb}{\faGithub}.
        }
    \label{fig:app-diffsettest}
\end{figure*}

\begin{figure*}[tb]
    \centering
    \includegraphics[width=0.9\linewidth,trim=0.5cm 0 0 0]{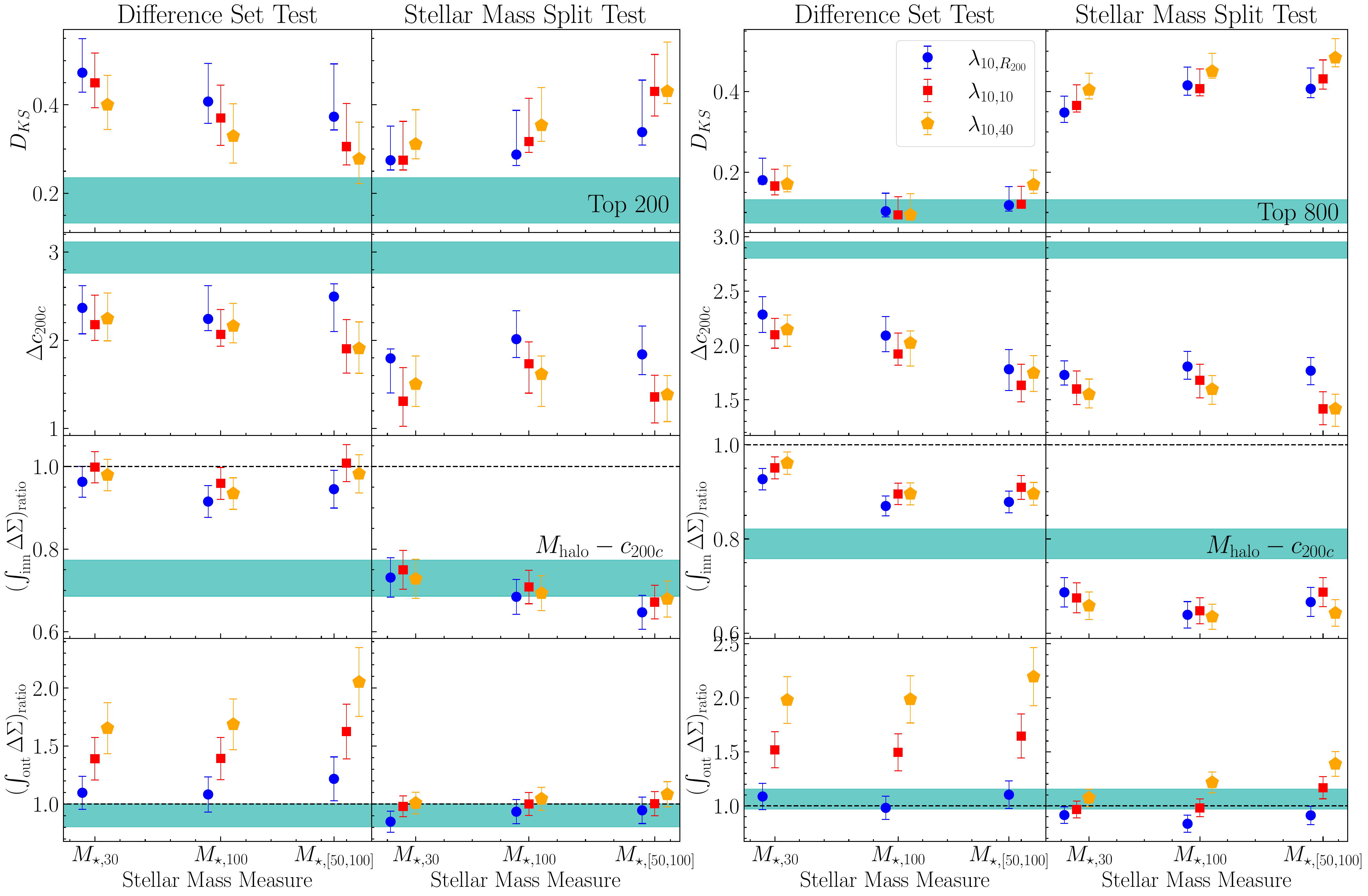}
    \caption{
        Similar to Fig. \ref{fig:assess}. Comparison of the performance of different set tests and stellar mass split tests at Top-N sample size $N=200,800$.
        The \texttt{Jupyter} notebook for reproducing this figure can be found here: \href{https://github.com/Xuchuyi/cenvssat/blob/main/FigureNotebooks/FigA2.ipynb}{\faGithub}.
        }
    \label{fig:app-assess}
\end{figure*}

    In Fig. \ref{fig:figkey}, we show the result for the difference-set test between outskirt stellar mass ($M_{\star,[50,100]}$) and satellite richness ($\lambda_{10, R_{200}}$) with a Top-N sample size of $N=600$. To further assess the performance of the difference-set test, we extend the analysis by adding cases at Top-$N$ sample sizes of $N=200, 400, 800$. These supplementary cases illustrate how test performance varies across different halo mass ranges.

    For the $N=200$ case, as noted below Fig. \ref{fig:topN}, $\lambda_{10,R_{200}}$ outperforms $M_{\star,[50,100]}$ as a halo mass proxy within this mass range. As a result, $\lambda$-only halos have higher average masses than $M_\star$-only halos. While $M_\star$-only halos still show earlier formation time and different 1-halo lensing profiles (Fig. \ref{fig:app-diffsettest}), the mass difference complicates efforts to attribute the observed lensing solely to secondary properties. As discussed in the main text, addressing this challenge requires modeling the hyperplane defined by the central galaxy's stellar mass, richness, halo mass, and a secondary halo property.

    At $N=400$ and $800$, the $M_\star$-only and $\lambda$-only halos have more similar mass distributions. $M_\star$-only halos exhibit a more centrally concentrated lensing signal, whereas the $\lambda$-only halos display a noticeable 'bump' feature in the intermediate radial range. Interestingly, the difference in the lensing profiles within the 2-halo term persists in the $N=400$ case but not in the $N=800$ case. As discussed in \S\ref{ssec:diff}, further investigation with larger-volume samples is necessary to determine whether this outer-region discrepancy arises from mass-dependent assembly bias or is merely an artifact of random effects.

    Then, Fig. \ref{fig:app-assess} extends the comparison of difference-set and stellar mass split tests (\S\ref{ssec:testchoice}) to $N=200$ and $N=800$, illustrating how test performance varies with halo mass range.

    For the sample ($N=200$) in the higher halo mass range, the large $D_{KS}$ value indicates that neither the difference-set test nor the stellar mass split test is an effective method for studying halo assembly history. The mass difference suppresses the concentration signal in the difference-set test while artificially enhancing it in the split test. Importantly, in the difference-set test, it is the low-$M_\star$ sample that exhibits higher halo mass, whereas in the stellar mass split test, it is the high-$M_\star$ sample that corresponds to higher halo mass. This suggests a hybrid approach may be needed when proxies perform differently.

    At $N=800$ (lower halo mass range), this case is similar to $N=600$. The difference-set test has a small $D_{KS}$ value while the stellar mass split test has a large $D_{KS}$ value, indicating the difference-set test is a better method to study halo assembly history at this halo mass range.

\section{Baryon Profiles of Halos in the difference-set test}
    \setcounter{figure}{0}

\begin{figure}[h]
    \centering
    \includegraphics[width=1.2\linewidth,trim=1.5cm 0 0 0]{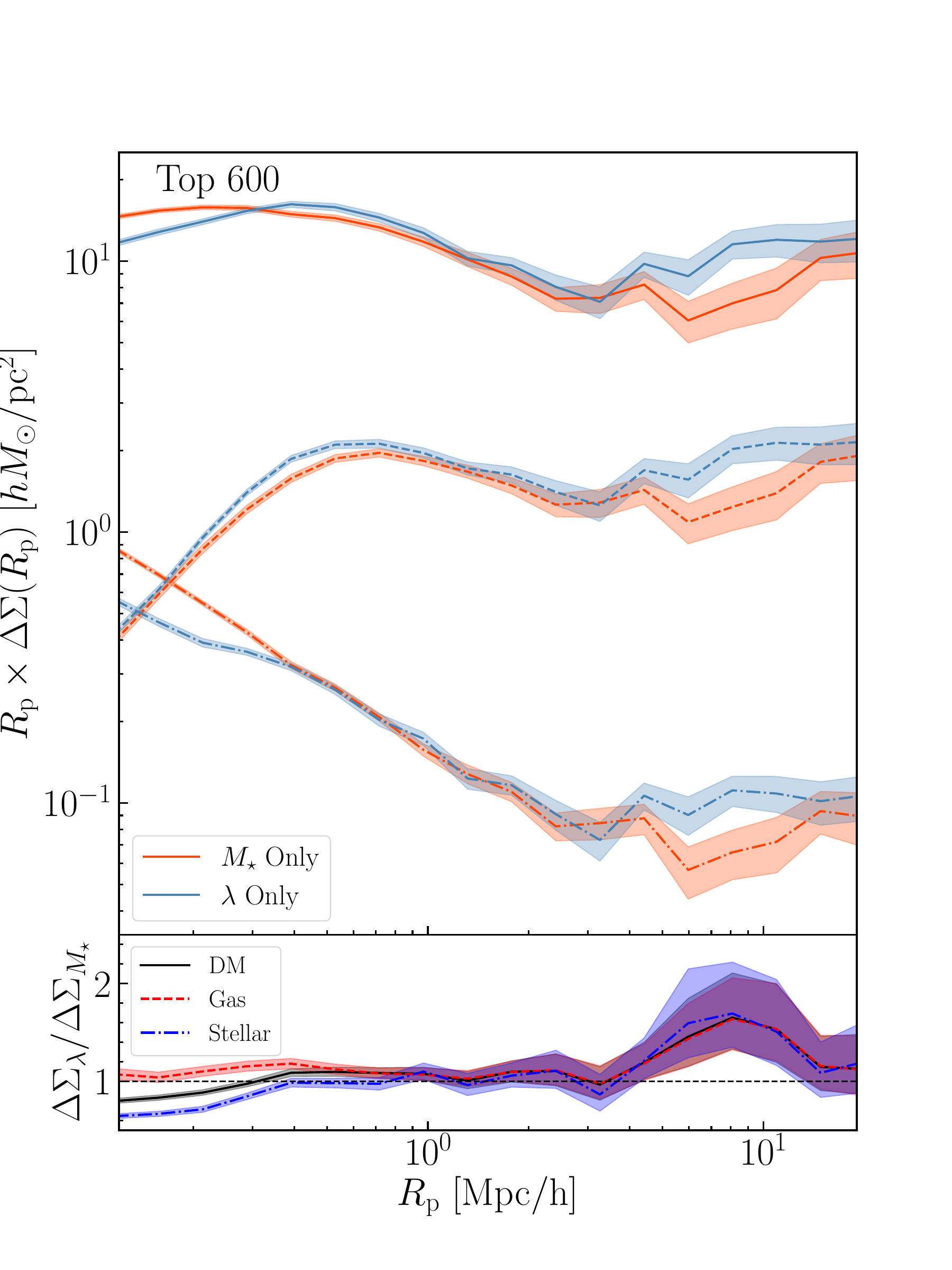}
    \caption{
        {\bf Upper panel:} The contribution to lensing profiles of Fig.\ref{fig:figkey} for dark matter (solid line), gas (dashed line), and stellar (dot-dashed line).
        {\bf Bottom panel:} The ratio of the lensing contribution between $M_\star$-only and $\lambda$-only for the three components.
        The lensing profiles show that dark matter dominates across the radius range ($R_{\rm p} \in [0.1,17]$ Mpc/h), indicating that the differences in lensing profiles between the samples are not due to the different distribution of baryonic components. 
        The \texttt{Jupyter} notebook for reproducing this figure can be found here: \href{https://github.com/Xuchuyi/cenvssat/blob/main/FigureNotebooks/FigB1.ipynb}{\faGithub}.
        }
    \label{fig:lenscomp}
\end{figure}

\begin{figure}[tb]
    \centering
    \includegraphics[width=0.9\linewidth,trim=1.5cm 0 0 0]{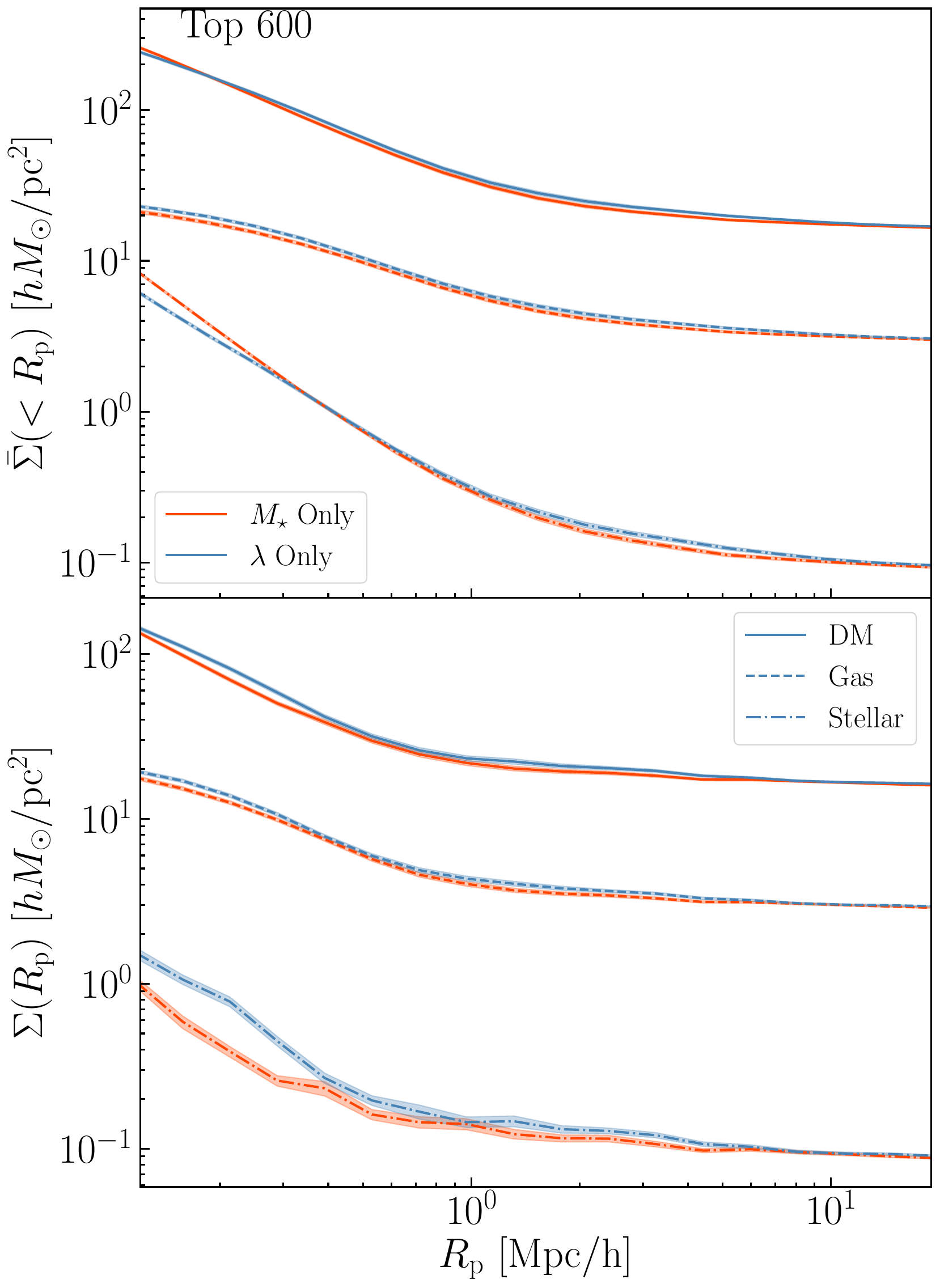}
    \caption{
        The surface density profile ($\Sigma(R_{\rm p})$ and $\bar{\Sigma}(<R_{\rm p})$) for dark matter, gas, and stars of the samples from the difference-set test in Fig. \ref{fig:figkey}.
        The \texttt{Jupyter} notebook for reproducing this figure can be found here: \href{https://github.com/Xuchuyi/cenvssat/blob/main/FigureNotebooks/FigB.ipynb}{\faGithub}.
        }
    \label{fig:app-baryonprof}
\end{figure}

    Since the baryonic effect primarily influences the baryonic components, besides the comparison between hydrodynamical and DMO simulations in \S\ref{ssec:baryon}, we also compare the contributions to the lensing profiles from different components for the two halo samples discussed earlier, as shown in Fig. \ref{fig:lenscomp}. We calculated the lensing contributions from dark matter, gas, and stars, as well as the corresponding ratio, $\Delta\Sigma_{\lambda}/\Delta\Sigma_{M_\star}$. Overall, dark matter dominates the lensing signal across $R_{\rm p}\in[0.1,17]$ Mpc/h, indicating that observed profile differences are not driven by baryonic components.

    Interestingly, the ratio of the lensing contributions from gas and stars, $\Delta\Sigma_{\lambda}/\Delta\Sigma_{M_\star}$, does not follow the same trend as the dark matter contribution in the 1-halo term. In the inner region of the halo ($R_{\rm p} \sim [0.1, 0.25]$ Mpc/h), the stellar contribution to the lensing profile is greater for the $M_\star$-only sample compared to the $\lambda$-only sample, while the gas contributions are similar between the two samples. However, in the outer region ($R_{\rm p} \sim [0.25, 1]$ Mpc/h), the gas contribution from the $\lambda$-only sample exceeds that of the $M_\star$-only sample, whereas the stellar contributions remain comparable in both samples. These findings suggest that halos with different assembly histories also exhibit distinct gas distributions, as previously noted in the literature (e.g., \citealt{Farahi2022ApJ}; \citealt{Marini2025A&A}).

    How can we interpret these differences? The difference in the excess surface density profile ($\Delta\Sigma(R_{\rm p})$) actually reflects both the magnitude and the shape of the surface density profile ($\Sigma(R_{\rm p})$ and $\bar{\Sigma}(<R_{\rm p})$). Therefore, we compare the $\Sigma(R_{\rm p})$ and $\bar{\Sigma}(<R_{\rm p})$ profiles to explore how the gas and stellar spatial distributions differ across the outskirts stellar mass - satellite richness plane. As Fig. \ref{fig:app-baryonprof} shows, we observe that the stellar component of the $M_\star$-only sample is more concentrated, with a steeper slope in the inner region. This result is consistent with the massive halo selection method we used. Regarding the gas contribution, although the gas in the $\lambda$-only halos is more concentrated in the inner region compared to the $M_\star$-only sample, the flatter slope of its profile leads to similar lensing contributions for both samples. Since the halo masses of both samples are similar, the average gas surface density in the halos is comparable, resulting in a steeper $\bar{\Sigma}(<R_{\rm p})$ profile and a higher lensing contribution of the $\lambda$-only sample in the outer region. 

\section{Choice of Richness}
\label{app:choiceRich}
\setcounter{figure}{0}

    \begin{figure}[tb]
    \centering
    \includegraphics[width=1.02\linewidth,trim=2cm 0 0 0]{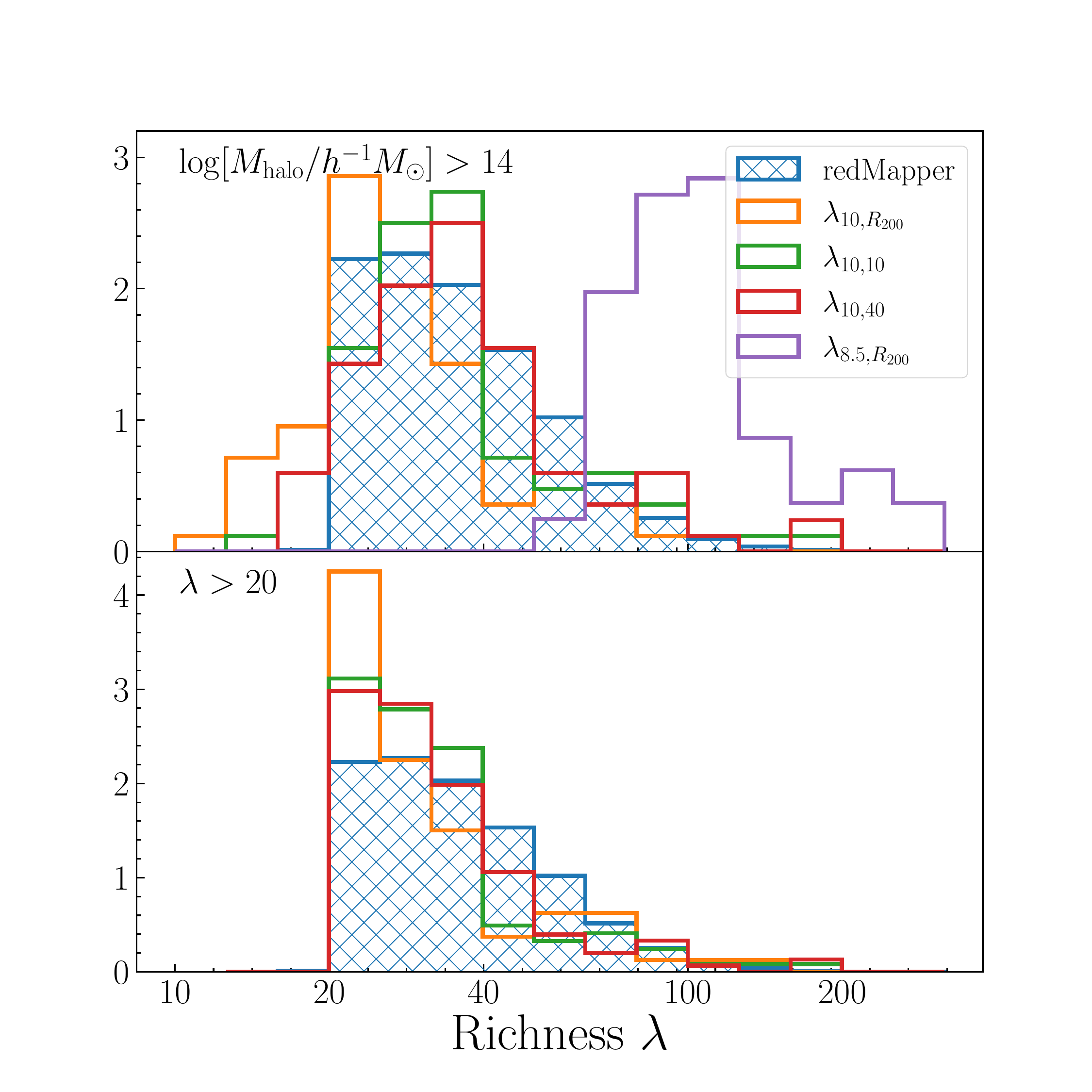}
    \caption{
        Comparison of the richness distribution of galaxy clusters from {\tt TNG300} sample in this work and SDSS DR8 redMapper catalog. We employ histograms for {\tt TNG300} clusters, along with a meshed blue histogram for the redMapper catalog, to distinguish richness definitions by color. We show 4 kinds of richness definitions here in \S\ref{ssec:richmea} based on stellar mass and location of satellites: $\lambda_{8.5,R_{200}}$ (purple), $\lambda_{10,R_{200}}$ (orange), $\lambda_{10,10}$ (green), and $\lambda_{10,40}$ (red).  
        \textbf{Upper Panel:} This panel shows the richness distribution of {\tt TNG300} clusters with $\log[M_{\rm halo}/h^{-1}M_\odot]>14$, which approximately corresponds to the halo mass range of the redMaPPer catalog, demonstrating a closer similarity in satellite counts with stellar mass larger than $10^{10}M_\odot$ to the richness definition in redMapper compared to those with stellar mass larger than $10^{8.5}M_\odot$, based on their values.
        \textbf{Bottom panel}: This panel shows the richness distribution of {\tt TNG300} clusters with satellite counts exceeding 20, revealing that richness defined within a cylindrical volume aligns more closely with the redMapper catalog than that defined within a spherical volume. 
        The \texttt{Jupyter} notebook for reproducing this figure can be found here: \href{https://github.com/Xuchuyi/cenvssat/blob/main/FigureNotebooks/FigC1.ipynb}{\faGithub}.
    }
    \label{fig:figC1}
    \end{figure}

    To check which definition of richness is more similar to the richness observed, we compare the richness distribution of our sample with that of the SDSS DR8 redMapper catalog (\citealt{Rykoff2014ApJ}), as shown in Fig.\ref{fig:figC1}. This catalog includes clusters with richness $\lambda\geq20$, which roughly correspond to massive halos with masses larger than $10^{14}~M_\odot$. We compare its richness distribution with that of halos with masses larger than $10^{14}~M_\odot$ in our sample, shown in the top panel of Fig.\ref{fig:figC1}. As anticipated, the richness distribution defined for satellites with stellar masses larger than $10^{10}~M_\odot$ roughly matches the range observed in the redMapper catalog. Moreover, the bottom panel further examines the richness distribution for halos with $\lambda>20$ in our sample, revealing that the richness definition that counts satellites within a cylindrical region (accounting for the projection effect) more closely resembles the distribution of the redMapper catalog than the definition that counts satellites within a spherical region. Based on these findings, we focus on the richness defined by a stellar mass threshold of $10^{10}~M_\odot$ in the analysis.


\bibliography{histrich}{}
\bibliographystyle{aasjournal}

\end{CJK*}
\end{document}